\documentclass[structabstract]{aa}
\usepackage{graphicx}
\usepackage{txfonts}

\begin{document}

   \title{Hysteresis of atmospheric parameters of 12 RR Lyrae stars\\
based on multichannel simultaneous Str\"omgren photometry\thanks{Original 
data (photometric indices) for each star are available in electronic form 
at the CDS via anonymous ftp to cdsarc.u-strasbg.fr (130.79.128.5) 
or via http://cdsweb.u-strasbg.fr/cgi-bin/qcat?J/A+A/}}

   \author{K.S. de Boer 
   \and G. Maintz 
 }

\titlerunning{Hysteresis of atmospheric parameters of 12 RR Lyrae stars}

\institute{Argelander Institut f\"ur Astronomie, Universit\"at Bonn, 
Auf dem H\"ugel 71, D-53121 Bonn, Germany.\\
deboer$|$gmaintz@astro.uni-bonn.de
}

\date{Received 11 January 2010 / Accepted 17 May 2010}

\abstract
{}
{RR Lyrae stars have been observed 
to improve the insight into the processes at work in their atmospheres.}
{Simultaneous Str\"omgren-photometry allows to obtain a rapid sequence of 
measurements in which photometric indices are unaffected by 
non-optimum observing conditions. 
The indices $y$, $b-y$, and $c_1$ are used with an established calibration 
to derive $T_{\rm eff}$ 
and to derive the gravity, $\log g_{\rm BJ}$ from the Balmer jump, 
throughout the pulsation cycle. 
By employing the equations for stellar structure, 
additional parameters can be derived.}
{Str\"omgren photometry and its calibration in terms of $T_{\rm eff}$ 
and $\log g$\ 
can be used to determine the run of $R$ and the atmosphere pulsation velocity. 
We find that the Balmer-line strengths are correlated with $T_{\rm eff}$ 
and that 
the strength of the Ca\,{\sc ii}\,K line correlates well with the radius 
of the star and thus the pulsation-dependent density of the atmosphere. 
The density in the stellar atmosphere fluctuates as indicated 
by the changes in the gravity $\log g_{\rm BJ}$, derived from $c_1$, 
between 2.3 and 4.5 dex. 
Also the Str\"omgren metal index, $m_1$, fluctuates. 
We find a disagreement between $\log g(T,L,M)$, 
the gravity calculated from $T_{\rm eff}$, $L$, and the mass $M$, 
and the gravity $\log g_{\rm BJ}$. 
This can be used to 
reassess the mass and the absolute magnitude of an individual star. 
The curves derived for the pulsational velocity $V_{\rm pul}$ differ from 
curves obtained from spectra needed to apply the Baade-Wesselink method; 
we think these differences are due to phase dependent differences in the 
optical depth levels sampled in continuum photometry and in spectroscopy. 
We find an atmospheric oscillation in these fundamental mode RR\,Lyrae stars 
of periodicity $P/7$.}
{Carefully conducted Str\"omgren-photometry allows to derive a large number 
of parameters for RR\,Lyrae stars. 
It provides a means of deriving masses and absolute magnitudes. 
When comparing photometry results with spectroscopic analyses 
it appears that optical depth effects affect all interpretations.}


\maketitle


\section{Introduction}

When lower mass stars evolve and come to core He burning, 
their properties can become those of the ``instability strip'' in the HRD. 
The atmospheric structure then reaches, 
as it were, a certain ``undecidedness'', 
causing rhythmic attempts to reach a stable state. 
The atmosphere cannot find its ``thermal equilibrium'' in 
the sense explored at a general level by Renzini et\,al. (1992) 
in the gedankenexperiment the ``gravothermal hysteresis cycle''. 

The rhythmic expansion and contraction of the atmosphere of 
pulsating RR\,Lyrae stars is caused by the $\kappa$-effect. 
The optical depth $\kappa$ in the layer in which He is ionized 
has a rhythmic variation: 
a higher level of ionization causes lower opacity, 
leading to a higher photon throughput, causing local cooling, 
which leads to recombination and thus to higher opacity, 
reduced radial energy transport so to increasing inner temperatures, 
coming full circle to increase ionization 
(see, e.g., Cox 1980; Gautschy \& Saio 1995). 
This cyclic behaviour produces hysteresis effects 
in the colour indices of the emergent stellar radiation 
as well as in various other observables (such as spectral line strengths). 
These all are based on, of course, hysteresis in 
surface gravity and effective temperature. 

Using photometry of RR Lyr stars, 
one can derive the variation in the parameters of the stellar surface. 
The first investigations of this kind were carried out by 
Oke \& Bonsack (1960), Oke et\,al. (1962), Oke (1966), and 
Danziger \& Oke (1967), who used spectral scanner data 
in comparison with model atmospheres to obtain $T_{\rm eff}$ and $\log g$ 
to calculate the change in radius. 
They noted that the changes in the atmosphere imply that 
one observes light from different atmospheric layers 
so that the radii thus derived might not be reliable. 

The information can also be obtained from Str\"omgren photometry. 
After the establishment of the Str\"omgren-photometric system and its early 
calibration (see, e.g., Breger 1974) this photometry was used 
by van Albada \& de Boer (1975) to derive the parameters 
$\Theta=5040/T_{\rm eff}$, $\log g$, and $R$ for all phases of the pulsation. 
McNamara \& Feltz (1977) and Siegel (1982) also used the Str\"omgren-system. 
Similar studies were performed by Lub (1977a, 1977b, 1979) 
with the Walraven photometric system. 

Only for a few RR\,Lyrae stars has the cycle of variation been followed 
in photometric detail in the Str\"omgren system. 
Thus only for a few of these stars is the variation in parameter values 
over the cycle accurately known. 
Photometry is normally performed sequentially 
in whichever wavelength bands selected. 
Sequentiality mandates good photometric conditions 
to obtain accurate colour indices 
for the derivation of reliable stellar parameters. 
Moreover, sequential measurements in the chosen bands are 
asynchronous among the bands. 
Performing measurements simultaneously in well selected photometric bands 
allows to obtain precise colour indices 
even in poor photometric conditions.  
For the research presented here, we used the 
Bonn University Simultaneous CAmera, {B\sc usca} (Reif et\,al. 1999). 
To transform the colour indices to astrophysical parameters, the 
calibration of 
the Str\"omgren system by Clem et\,al. (2004) has been used. 
{B\sc usca} also allows a rapid succession of measurements 
leading to a better coverage of fine structure in light curves. 

The Baade-Wesselink method (Baade 1926, Wesselink 1946) requires the 
measurement of the radial velocity variation to produce 
a full characterization of a pulsating star. 
Following van Albada \& de Boer (1975) we will again show that, 
with accurate photometry, $R$ can be derived 
(rather, the apparent angular extent, 
which indicates the radius when the distance to the star is known) 
and the changes in $R$ 
can be used to calculate the run of atmospheric velocities $V_{\rm pul}$ 
through the pulsational cycle. 
Str\"omgren photometry allows, in principle, 
to derive astrophysical parameters 
more accurately than possible from the widely used Johnson photometry.

\begin{figure}
\resizebox{\hsize}{!}{\includegraphics{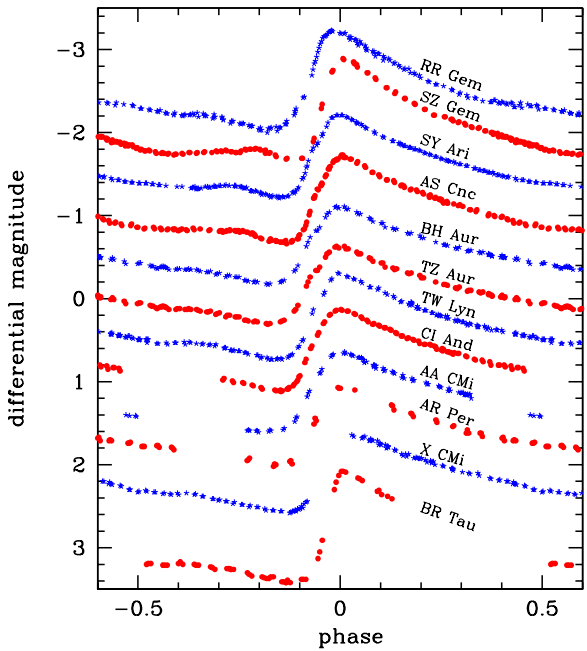}}
\caption[]{Lightcurves of the RR\,Lyrae stars in $y$. 
Data from subsequent cycles have been combined 
(see Table\,\ref{tabobs} for observing dates). 
The curves have been shifted arbitrarily in $y$ to fit them into one panel. 
For the stars lower in the figure, the lightcurve coverage is incomplete.}
\label{flightcurves}
\end{figure}

\begin{figure}
\resizebox{\hsize}{!}{\includegraphics{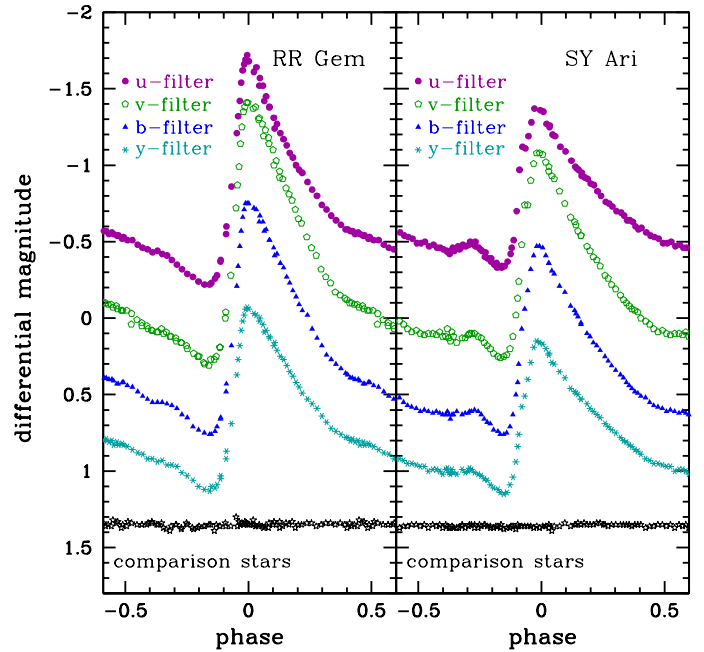}}
\caption[Photometry without beta]{Lightcurves of 
\object{RR Gem} and \object{SY Ari} 
observed with {B\sc usca} in the Str\"omgren-system. 
Note the different brightness amplitude of the two stars.  
At the bottom, the instrumental brightness of the mean of two 
comparison stars (selected in the field of each variable) is shown.
}
\label{photn}
\end{figure}

\section{Observations and data reduction}
\label{secobs}

The observations were carried out in three sessions at the 
Calar Alto Observatory of the Deutsch-Spanisches Astronomie Zentrum (DSAZ) 
in southern Spain in three successive winter seasons, 
between 4 and 9 Jan.~2005, between 16 and 19 Dec.~2005, 
and between 26 and 28 Nov.~2006. 

The times of maximum brightness 
within the observing sessions were calculated for each star from 
the period $P$ and a reference epoch as 
$V_{\rm max} = {\rm JD}({\rm ref.epoch})) \times (nP)$ with appropriate $n$. 
This allowed optimal phase-coverage planning of the photometry. 
In the various figures to follow, 
the data are plotted against phase $\Phi$ based on 
the observed maximum. 
For a few stars, the maximum could not be measured 
so the phase was determined from early epoch maxima. 

Our programme consisted of observing 18 relatively bright RR\,Lyrae 
(for accurate positions and additional data, see Maintz 2005). 
Of 12 of these, (almost) complete lightcurves were obtained 
(coverage $>$70\%) as given in Table\,\ref{tabobs} 
and Fig.\,\ref{flightcurves} (and Fig.\,\ref{photn}).
However, for X\,CMi we missed the maximum. 
Of 6 stars (GM\,And, OV\,And, GI\,Gem, TY\,Cam, TT\,Lyn, CZ\,Lac), 
only smaller portions (15 to 57\%) of the light curve could be obtained; 
these stars are not discussed further. 

\begin{figure}
\resizebox{\hsize}{!}{\includegraphics{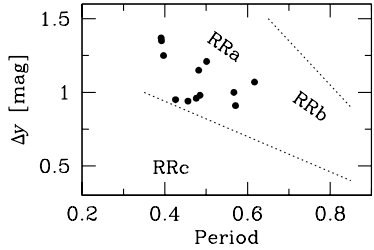}}
\caption[]{Plot of the amplitude in $y$ versus period of our stars  
showing that they are fundamental mode RRa-type pulsators. 
The band between the dotted lines marks the zone of RRa and RRb pulsators 
(see Ledoux \& Walraven 1958).
}
\label{rrtype}
\end{figure}

The stars selected are fundamental mode pulsators (see Fig.\,\ref{rrtype}). 
They generally have lightcurve bumps. 

\begin{table*}
\caption[Table of dates of observations]{The stars with basic information, the J.D. of the observations, and the phase ranges covered}
\begin{tabular}{lrrrrrr}
\hline
\hline
Star &  \object{CI And} & \object{SY Ari}  & \object{AR Per} & \object{BR Tau}   & \object{BH Aur}  &   \object{TZ Aur}    \\
\hline
RA  &            01 55 08.29 & 02 17 34.04  & 04 17 17.19   & 04 34 42.89 & 05 12 04.26  & 07 11 35.01  \\
DEC &            43 45 56.47 & 21 42 59.26  & 47 24 00.63   & 21 46 21.72 & 33 57 46.95  & 40 46 37.13  \\
mean $y$ ; \ $E(B-V)$ \ [mag] & 11.86 ; 0.22&12.21 ; 0.25&9.40 ; 0.73&11.96 ; 0.31 & 11.43 ; 0.19 & 11.43 ; 0.19 \\
distance [pc] ; \ [M/H] & 1741 ; $-$0.83 & 2100 ; $-1.4:$\tablefootmark{a} & 612 ; $-$0.43 & 1835 ; $-0.7:$\tablefootmark{a} & 1400 ; +0.14 &  1500 ; $-$0.79 \\
period [d]    & 0.484728     & 0.5666815    & 0.425551      & 0.3905928   & 0.456089     & 0.39167479   \\
exp. time [s] $v,b,y$; $u,b,y$ & 40; 40 & 70; 70 & 20; 60 & 70; 210 & 70; 210 & 30; 30\\ 
start of observations & & & & & & \\
\hspace*{3mm}at JD = 2453370. & +5.2914      & +5.2947     &              &              &              & +5.6153      \\
phase covered & 0.82 to 1.26 & 0.66 to 1.05 &              &              &              & 0.67 to 0.95 \\
              &  +7.4215     & +7.2612      &              &              &              & +6.4747      \\
              & 1.21 to 1.43 & 0.13 to 0.65 &              &              &              & 0.86 to 1.60 \\
              &              & +8.3353      &              &              &              & +7.5535      \\
              &              & 1.03 to 1.16 &              &              &              & 0.62 to 0.97 \\ 
              &              & +9.2568      &              &              &              & +8.3106      \\
              &              & 0.65 to 0.91 &              &              &              & 0.55 to .89  \\
\hspace*{3mm}at JD = 2453720. & +1.32        &              &              & +3.2969      & +1.2980      &              \\
phase covered & 0.66 to 0.85 &              &              & 0.53 to 1.14 & 0.52 to 1.39 &              \\
              &              &              &              &              & +2.5868      &              \\
              &              &              &              &              & 0.34 to 0.54 &              \\
\hspace*{3mm}at JD = 2454060. &              &              & +7.2981      & +7.5013      &              &              \\
phase covered &              &              & 0.75 to 1.63 & 0.73 to 0.87 &              &              \\
\\[-6pt]
\hline
\hline
Star &   \object{AA CMi}   & \object{RR Gem}  &  \object{X CMi} & \object{TW Lyn}  & \object{SZ Gem}      &  \object{AS Cnc}   \\
\hline
RA  &           07 17 19.17  & 07 21 33.53  & 07 21 44.62  & 07 45 06.29  & 07 53 43.45  & 08 25 42.11 \\
DEC &           01 43 40.06  & 30 52 59.45  & 02 21 26.30  & 43 06 41.56  & 19 16 23.93  & 25 43 08.80 \\
mean $y$ ; $E(B-V)$ [mag]&11.13 ; 0.16&11.03 ; 0.22&12.23 ; 0.35&11.91 ; 0.16 & 11.45 ; 0.13 & 12.58 ; 0.16\\
distance [pc] ; \ [M/H]  & 1220 ; $-$0.15 & 1260 ; $-$0.29 & 1766 ; $-$0.71 & 1549 ; $-$0.66 & 1601 ; $-$1.46 & 2324 ; $-$1.89 \\
period [d]     & 0.476327    & 0.397292     & 0.57138      & 0.481862     & 0.5011270    & 0.61752     \\
exp. time [s] $v,b,y$; $u,b,y$ & 40; 40 / 20; 60 & 20; 20 & 70; 70 & 40; 40 & 20; 20 & 40; 40 \\
start of observations & & & & & & \\
\hspace*{3mm}at JD = 2453370. &              & +5.5257      & +8.6759      & +6.5937      & +5.5385      & +7.5695      \\
phase covered &              & 0.98 to 1.49 & 1.08 to 1.17 & 1.16 to 1.51 & 0.39 to 0.77 & 0.73 to 1.05 \\
              &              & +6.4776      & +9.3641      & +7.2652       & +6.4800      & +8.4688      \\
              &              & 0.38 to 1.10 & 0.29 to 0.93 & 0.55 to 0.76 & 1.26 to 1.48 & 1.18 to 1.63 \\
              &              & +8.5121      &              & +8.3421      & +7.6349      & +9.4268      \\
              &              & 0.50 to 0.88 &              & 0.79 to 1.25 & 0.57 to 0.84 & 0.73 to 1.28 \\
              &              &              &              & +9.2625      &              &              \\
              &              &              &              & 0.70 to 0.76    &              &              \\
\hspace*{3mm}at JD = 2453720. & +2.4285   &              & +1.5757      &             &  +1.5348     & +1.4625      \\
phase covered & 0.77 to 1.09 &              & 1.21 to 1.32 &              & 0.82 to 1.28 & 0.59 to 0.69 \\
\hspace*{3mm}at JD = 2454060. & +6.6066      &              & +6.6110      &              &              &              \\
phase covered & 1.02 to 1.32 &              & 1.07 to 1.28 &              &              &              \\
\hline
\hline
\end{tabular}

\vspace*{1mm}
\tablefoottext{a} [M/H] (in logarithmic units) estimated from our $m_1$ 
(see Sect.\,\ref{colourloops}).
\label{tabobs}
\end{table*}

{B\sc usca} (Reif et\,al. 1999) is operated at the 2.2m telescope of the 
Calar Alto Observatory. 
{B\sc usca} splits the telescope beam above the Cassegrain focus 
via dichroic beam splitters in 4 wavelength channels 
between 3200 and 9000 \AA, with edges at 4300, 5400 and 7300 \AA. 
Each channel is equipped with a (4k)$^2$ CCD. 
In each channel, a filter wheel allows the placement of 
appropriate filters. 

For the measurements we used the Str\"omgren $u,v,b,y$-system, 
extended by the Cousins $I$ band\footnote{Since the Cousins $I$ 
is about 10 times wider than the bands of the Str\"omgren-system, 
the $I$ measurements were mostly overexposed and have not been used 
for this paper. 
For a few stars we also made some measurements in the H-Balmer filters 
H$\beta$W (wide) and H$\beta$N (narrow), 
which are in the same channel as the Str\"omgren-$b$ filter. 
Since the H$\beta$N (narrow) filter is yet much narrower than the 
Str\"omgren-filters, exposure times would have to be excessively long and 
coverage of the light curves in these filters poor 
so these data have not been used either.}. 
The $y$ band is close to the cut-off of a {B\sc usca}-dichroic filter; 
however, the transformation of $y$ into the standard system 
could be performed without problems. 
Because of the proximity of the wavelengths of $u$ and $v$, 
their filters are in the same {B\sc usca} channel. 
Since the $u$ and $v$ bands could not be measured simultaneously, 
the photometry alternated between the simultaneous exposures in 
$y,b,v,(I)$ and $y,b,u,(I)$. 
For two stars, light curves in the four Str\"omgren-bands 
are shown in Fig.\,\ref{photn}. 
More data are shown in Maintz (2008a). 

To calibrate the data, 
the instrumental brightness of each variable was obtained by comparison 
with two non-variable stars in the field. 
Where possible, one star with a brightness near the $y_{\rm max}$ 
of the RR\,Lyrae star was used 
and another of brightness close to $y_{\rm min}$ of the variable. 
(In the data a known suspected variable was quantified; see Maintz 2008b.) 
The airmass correction was obtained from these comparison stars. 

To secure the absolute calibration, 
we observed appropriate stars from the list of Perry et\,al.\,(1987) 
at moments in the light cycle of the variable 
where the changes in brightness were small or monotonous. 
The instrumental magnitudes of the (airmass-corrected) photometry 
were related to the true magnitudes of the calibration stars. 
This permitted the calibration of the RR\,Lyrae star photometry. 
First, the RR\,Lyrae $y$ measurements were calibrated. 
Since the instrumental $(b-y)_{\rm inst}$ from {B\sc usca} is more accurate 
than $b_{\rm cal}$-$y_{\rm cal}$ (the same is true for the other indices), 
the calibrated $b-y$ and $v-b$ were obtained from the instrumental index 
and that of the calibration stars. 
The sequentially obtained colour indices $u-y$ and $v-y$ were then used 
to interpolate in time to obtain $u-v$, which was calibrated. 
Using these indices, 
$c_1= (u-v)-(v-b)$ and $m_1 = (v-b) - (b-y)$ were calculated. 

The RR\,Lyrae star CCD photometry was performed with equal exposure times 
throughout the cycle, 
of 20\,s for the brightest and 70\,s for the faintest stars. 
Exposure times are given in Table\,\ref{tabobs}. 
For three stars the exposure times in the $u,b,y$ measurements were 
$3\times$ those of the $v,b,y$ measurements (see Table\,\ref{tabobs}). 
The (simultaneous) reading-out of the CCDs took about 2~min. 
Thus the CCD exposures followed each other quickly, 
within 2.5\,min for the brighter stars and 3.5\,min for the fainter ones. 
The short exposure times did not lead to large time smearing, 
not even when the star changed quickly in brightness. 
In most cases, two stars were observed alternatingly in rapid succession 
to achieve good light curve coverage for as many stars as possible. 
This alternation explains why most light curves presented have data intervals 
longer than 3.5~min. 

Basic data ($d$, $E(B-V)$, [M/H]) were taken from Beers et\,al. (2000), 
Fernley et\,al. (1998), and Layden (1994), in part taking averages. 
Some of the stars are known to be behind gas with extinction 
and we corrected for that using expressions given by Clem et\,al. (2004). 
For four stars we calculated $d$ from $M_V$ based on [Fe/H] 
using the formula of Fernley et\,al. (1998). 
The adopted distances are given in Table\,\ref{tabobs}. 

\section{Colour-index loops}
\label{colourloops}

\begin{figure}
\resizebox{\hsize}{!}{\includegraphics{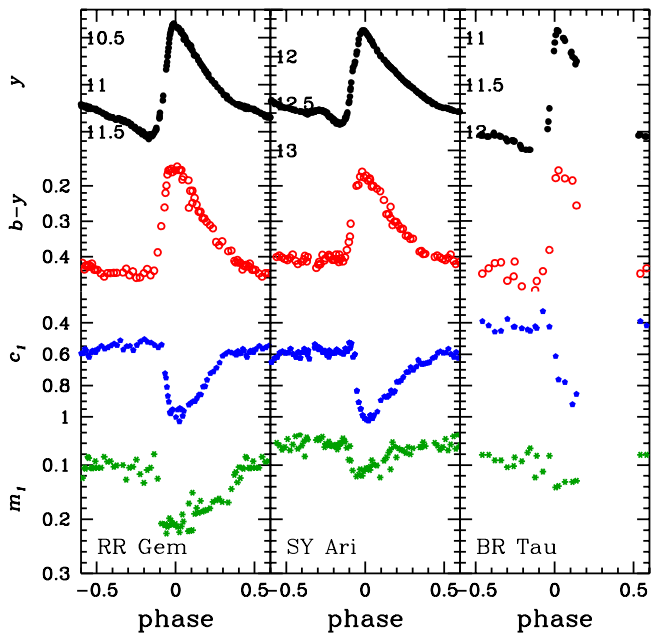}}
\caption[Photometry Stroemgren indices]{Lightcurves in the 
calibrated Str\"omgren-indices of the stars \object{RR~Gem}, \object{SY Ari}, 
and \object{BR Tau}. 
The phase coverage of the photometry 
can be reconstructed from the information given in Table\,\ref{tabobs} 
(for \object{BR~Tau} the coverage is only 70\%) 
and can be recognised in Fig.\,\ref{flightcurves}.}
\label{photstrom}
\end{figure}

\begin{figure*}
\resizebox{\hsize}{!}{
\includegraphics{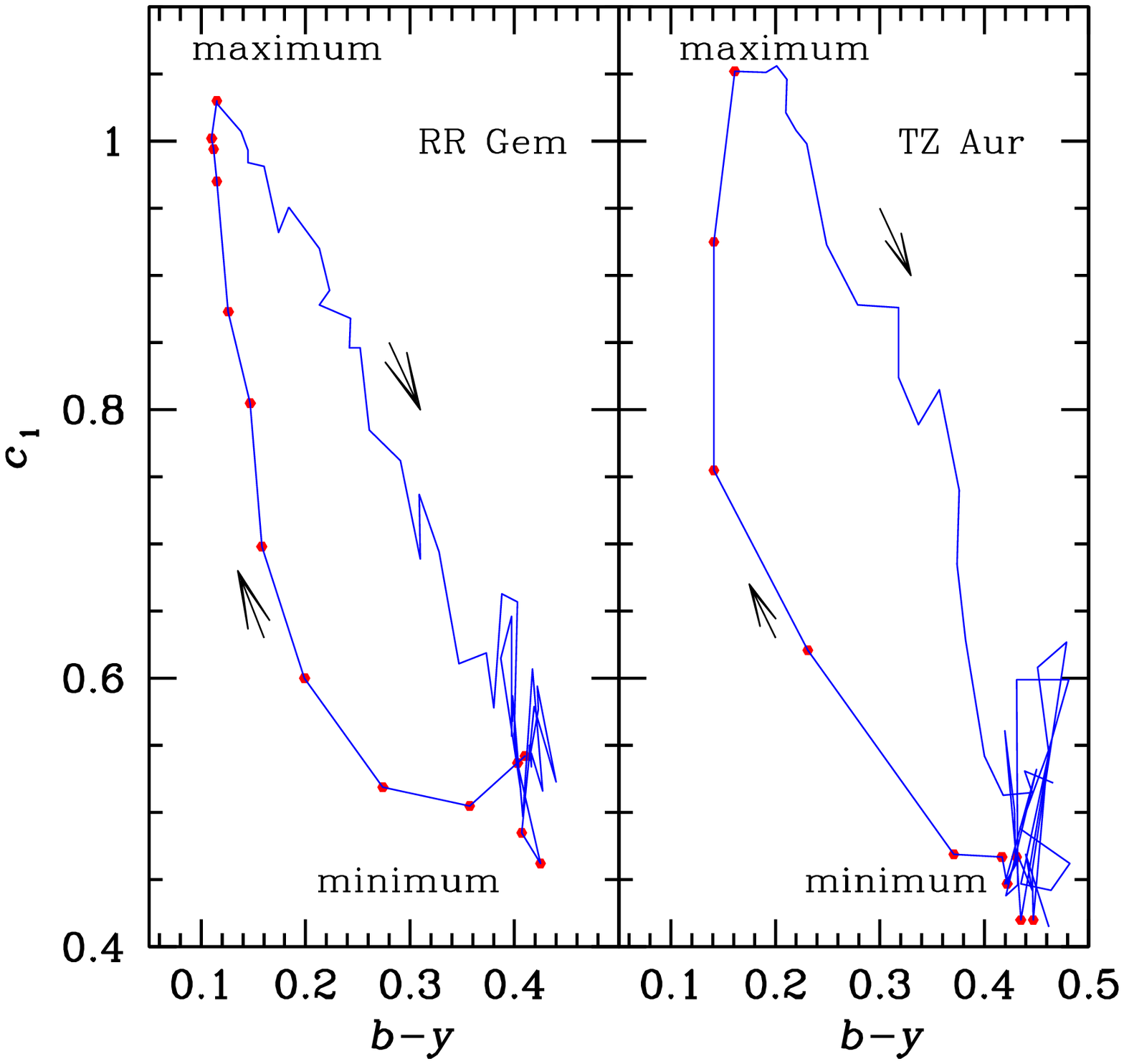}
\includegraphics{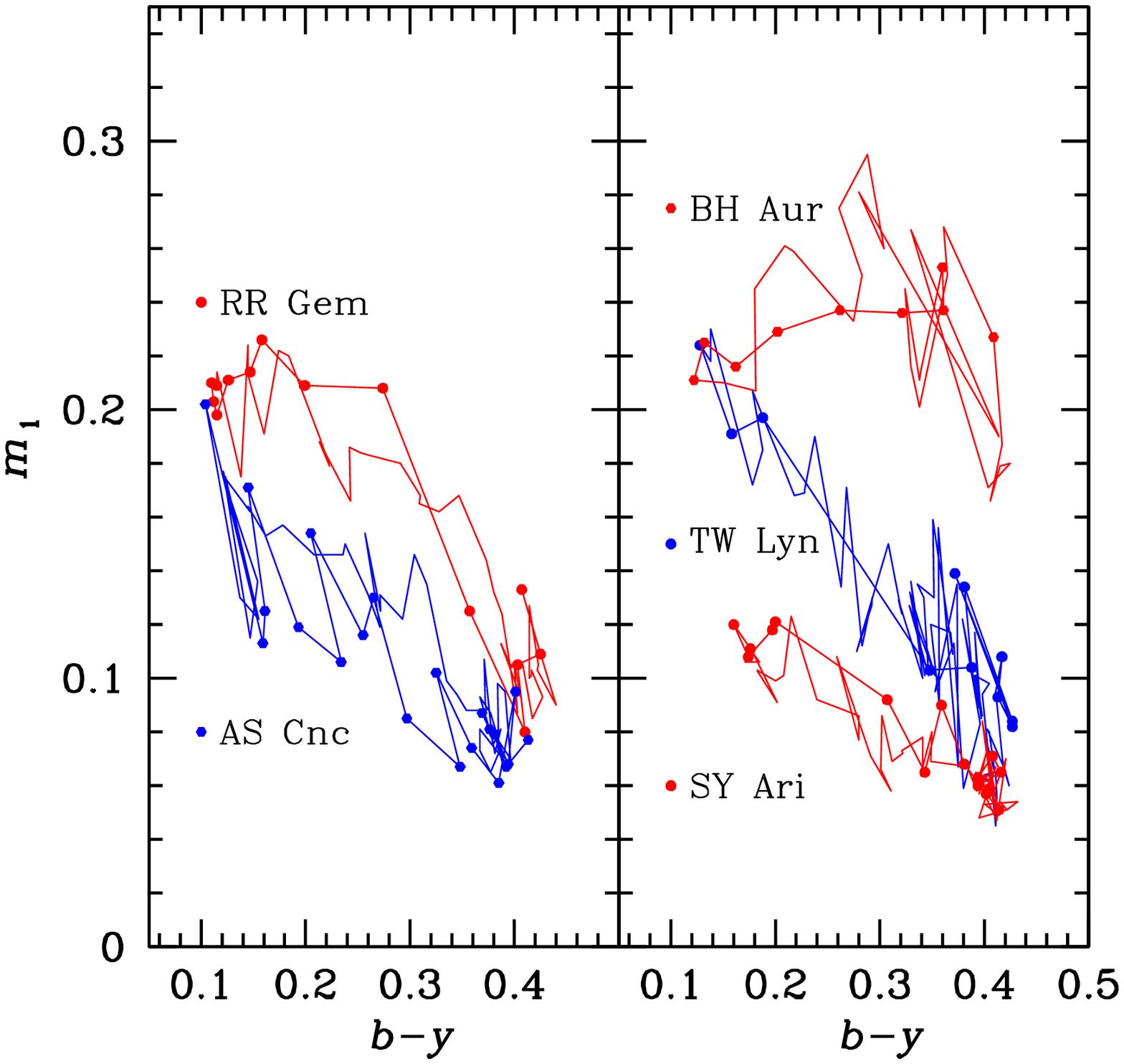}
}
\caption[colour loops]{
Colour loops in $b-y,c_1$ (at left) 
and in $b-y,m_1$ (at right) of several stars. 
The data points of the brightening phase are marked with~$\bullet$.\\
{\bf Left:} 
Clear hysteresis-like behaviour in $b-y,c_1$ is shown for 
\object{RR Gem} and \object{TZ Aur}. 
Arrows mark the direction of time. 
At phases between 0.3 and 0.85 (see Fig.\,\ref{photstrom}), 
the colour index $c_1$ seems to just scatter. 
The $b-y,c_1$ combination represents mostly $T_{\rm eff}$ 
but also Balmer excitation.\\
{\bf Right:} 
Colour loops in $b-y,m_1$ 
(shown for \object{RR Gem}, \object{AS Cnc}, \object{SY Aur}, \object{TW Lyn}, 
and SZ~Gem) marginally exhibit hysteresis-like behaviour. 
The index $m_1$ generally represents metal content [M/H], 
its change in RR\,Lyrae stars is mostly due to changing $T_{\rm eff}$. 
}
\label{colloops}
\end{figure*}

For three stars the run through the cycle of the derived 
absolutely calibrated colour indices is shown in Fig.\,\ref{photstrom}. 
The $b-y$ versus $c_1$ diagram for two typical stars from our sample, 
RR Gem and TZ Aur (Fig.\,\ref{colloops}, left panel), 
shows clear colour index loops. 
The index $m_1$ varies in line with $c_1$ (Fig.\ref{photstrom}). 
Figure\,\ref{colloops} (right panel) 
shows the very moderate $m_1$ loops in a colour-colour plot. 

The ($c_1$,$b-y$) curves show the interplay 
of the surface parameters $\log g$ and $T_{\rm eff}$ during the cycle. 
The loop is essentially due to $c_1$, 
which represents (above $T_{\rm eff} \simeq 6000$\,K) 
the strength of the Balmer jump. 
At low $T_{\rm eff}$, the Balmer level is hardly populated. 
When the temperature increases (bluer $b-y$), 
$T_{\rm eff}$ and $n_e$ (and thus $\log g$) of the atmosphere gas increase 
and (collisional) excitation of hydrogen becomes noticeable. 
When the temperature decreases the population of the Balmer level decreases 
as well. 
In RR\,Lyrae star atmospheres, the ionisation of hydrogen hardly plays a role 
(except for an increase in $n_e$) because the highest temperatures 
in the pulsational cycle are $T_{\rm eff} <9000$~K.

It is evident from the data that 
the highest temperatures appear when the star is close to maximum light. 
During most of the period, the measured colour indices 
indicate low temperatures (large $b-y$, small $c_1$), 
as visible in the clump of data points in the lower-right corner 
of the left-side panels of Fig.\,\ref{colloops}. 
The brightening leg and the dimming leg of the colour curves 
do not overlap. 
During brightening, $b-y$ changes faster than $c_1$, 
which is indicative of a faster rise in $T_{\rm eff}$ than in 
the excitation to the Balmer level. 
In the dimming branch the converse is found (see Fig.\,\ref{colloops}, left).  
Thus the colour curve exhibits hysteresis. 

As for the loops in $m_1$ (Fig.\,\ref{colloops} right), 
these are quite moderate and do not show much 
in the sense of hysteresis. 
The change in $m_1$ over the cycle is on average 0.1 mag 
($m_1$ is largest at maximum light). 
For our stars, $m_1$ is larger for stars known to have larger [M/H] 
(see also the calibration for red giant stars by Hilker 2000), while 
for stars with [M/H] $<-1$\,dex $m_1$ is only marginally metal dependent. 
At the $T_{\rm eff}$ of RR\,Lyrae stars, 
cycle variations in $m_1$ related to the [M/H] are small. 
Nevertheless, using these trends, we estimated [M/H] for the two stars 
for which these values were unavailable from other sources 
(see Table\,\ref{tabobs}). 
The variation in $m_1$ over the cycle 
must be caused by changes in temperature and electron density, 
which both influence the level of ionisation and excitation of ions. 

\section{Behaviour of atmospheric parameters}

\subsection{Deriving parameters in the stationary case}

The parameters $T_{\rm eff}$ and $\log g$ were determined from our photometry 
using the conversion grids given by Clem et\,al. (2004). 
These grids are based on modelled spectral energy distributions and 
are provided in steps of 0.5 dex of metal content [M/H]. 
Clem et\,al.\ made extensive calibrations of the Str\"omgren system 
against the parameters of non-variable stars.
They show that individual Population\,II star colours 
can be matched with models to within $\simeq$200\,K and $\simeq$0.1 dex, 
showing what is the ultimate uncertainty of a match. 
For Population\,I stars, the effect 
of small variations in metal content (0.3 dex) on the colours, 
and thus on the derived $T_{\rm eff}$ and $\log g$, is small. 
The dependence of colours on metal content is significant only 
for $T_{\rm eff} < 6000$\,K. 
For each star we used the grid for which [M/H] was closest to 
the metal content of each star. 

The parameters $T_{\rm eff}$ and $\log g$ can be used to derive 
additional parameters describing the behaviour of the stellar atmosphere. 
The conditions in a regularly pulsating stars 
are in ``quasi-hydrostatic equilibrium'' (Cox 1980; Ch.\,8.3), 
meaning that the use of a calibration based on stable stars is justified. 

However, when a RR\,Lyrae star brightens from $V_{\rm min}$ to $V_{\rm max}$ 
within approximately one hour, 
the atmosphere is probably not in quasi-hydrostatic equilibrium. 
Doubling of Ca and H lines has been seen (e.g., Struve 1947, Sanford 1949) 
and H lines have been seen in emission (e.g., Preston \& Paczynski 1964) 
in that part of the RR\,Lyrae cycle, 
which according to Abt (1959) originate from dynamic effects. 
Since no models exist that take these effects for photometry into account 
there is no other option than to use static models. 
Furthermore, the light we detect comes from the photosphere, i.e., 
the layer in which, at the particular wavelength observed, 
the optical depth is $\tau$ $\simeq$ $0.7$. 
This means that in the course of the pulsational cycle it is not necessarily 
the same gas from which the light detected emanates. 
We return to these aspects later (Sect.\,\ref{sectau}). 

The luminosity of a star is related to the surface parameters 
$R$ and $T_{\rm eff}$ through the familiar equation 
$L = 4\pi R^2\cdot \sigma T_{\rm eff}^4$, 
which can be rewritten in the relative logarithmic form 
\begin{equation}
\log{\Bigl(\frac{L}{L_\odot}\Bigr)}  = 
2 \log{\Bigl(\frac{R}{R_\odot}\Bigr)} + 
4 \log{\Bigl(\frac{T}{T_\odot}\Bigr)} \ \ \ .
\label{elogLRT}
\end{equation}

The surface gravity of a star is defined as $g = {\rm G}\ {M}/{R^2}$. 
This expression can be transformed into a logarithmic one 
relative to solar values given by 
\begin{equation} 
\log{\Bigl(\frac{g}{g_\odot}\Bigr)}  = 
\log{\Bigl(\frac{M}{M_{\odot}}\Bigr)} 
- 2 \log{\Bigl(\frac{R}{R_\odot}\Bigr)} \nonumber \ \ \ .
\label{elogMRg}
\end{equation}

Using Eqs.\,\ref{elogLRT} and \ref{elogMRg}, one can then eliminate 
the radius so that the combined equation has 
the stellar mass, temperature, gravity, and luminosity as variables, 
\begin{equation}
\log{ \Bigl(\frac{L}{L_\odot}\Bigr)} + \log g = 
 4\log{T_{\rm eff}} + \log{ \Bigl(\frac{M}{M_{\odot}}\Bigr)} -10.68 \ \ \ ,
\label{elogMTgL}
\end{equation}
where $-10.68= (\log T_{\rm eff} -\log g)_{\odot}$ in cgs units. 
This equation can be used to calculate $\log g$.

\subsection{Deriving parameters in the pulsating case}
\label{parpulsating}

In a pulsating atmosphere the parameters $T_{\rm eff}$, $\log g$, and $R$ 
vary continuously 
(note that no time subscripts are used in this paper). 

\noindent
$\bullet$ $T_{\rm eff}$, $L$.\hspace*{2mm}
The time-dependent values $L$ and $T_{\rm eff}$ can be derived from the 
photometry if one knows the stars distance. 
$T_{\rm eff}$ is simply inferred from a calibrated $b-y$ 
(using the grid of Clem et\,al. 2004). 
$L$ has to be determined from $y$, 
the integral over the spectral energy distribution 
and the distance $d$ of the star. 

As for $L$, an RR\,Lyrae star has the convenient property that 
the maximum of its spectral energy distribution (in $B_{\lambda}$) 
lies close to the middle of the visual wavelength band, 
near Johnson $V$ or Str\"omgren $y$. 
This is true during the entire cycle: 
$T_{\rm eff}$ actually varies only between roughly 5000 and 9000\,K.
This means that 
the bolometric correction is neither large nor varies a lot. 
We have performed the bolometric correction 
using the classic values of Schmidt-Kaler (1982). 
We fitted a quadratic equation 
in the temperature range relevant for RR\,Lyrae stars. 

Distances were adopted as described in Sect.\,\ref{secobs}. 
If the distance $d$ of a star were off by 20\%, 
this would result in an error of 0.08 in $\log L$.

Results for $L$ and $T_{\rm eff}$ are shown in Fig.\,\ref{tefflloop}. 
 
\noindent
$\bullet$ $R$.\hspace*{2mm}
The radius can be calculated from Eq.\,\ref{elogLRT}. 
For examples of radius change see Fig.\,\ref{parcurve}. 

\noindent
$\bullet$ $\log g$, $\log g_{\rm BJ}$, $\log g_{\rm eff}$.\hspace*{2mm}
The surface gravity $g$ can be calculated from Eq.\,\ref{elogMTgL} 
if one knows the luminosity of the star (discussed above) 
and its mass (see below). 
This gravity is 
like an overall equilibrium gravity and is henceforth called $g(T,L,M)$. 

There is a further aspect affecting the gravity. 
In a pulsating atmosphere, the actual surface gravity 
($g(T,L,M)$ at the level $\tau \simeq 0.7$) is modified 
due to acceleration of the atmosphere during the pulsation cycle 
with respect to the ``normal'' gravity. 
Thus, $g$ is modified by an acceleration term, $d^2R/dt^2$. 
The total gravity is called the ``effective'' gravity 
\begin{equation}
g_{\rm eff} = g(T,L,M) + \frac{d^2R}{dt^2} \ \ \ ,
\label{eloggeff}
\end{equation}
indicating the vertical gravitational force in the atmosphere. 

The time-dependent Str\"omgren photometry measures 
the actual condition of a stellar atmosphere. 
The gravity derived from $b-y,c_1$ is a function of gas density, 
visible in the excitation of the Balmer level 
and in $c_1$ representing the Balmer jump. 
We will call this gravity the {\bf ``Balmer-jump'' gravity}, 
$\log g_{\rm BJ}$. 
This gravity is independent of $\log g_{\rm eff}$ and instead 
represents the pressure of the gas sampled, i.e., $P_{\rm gas}=f(T,\rho)$.

We will return to the values derived for $\log g(T,L,M)$ 
and $\log g_{\rm BJ}$ in Sect.\,\ref{secggeff}.

\noindent
$\bullet$ $M_V$, $M$.\hspace*{2mm}
RR\,Lyrae star distances have thus far been derived 
using a reference value for $M_V$. 
This value of $M_V$ does not apply to all RR\,Lyrae stars 
since horizontal-branch (HB) stars 
evolve over more than $\Delta \log L = 0.1$ on the HB. 
If, e.g., $M_V$ of a star were to differ by 0.25 mag from the 
reference value adopted, $L$ would be off by 0.10. 

The mass of RR\,Lyrae stars is in the range from $0.6$ to $0.8$\,M$_{\odot}$.
We adopted a mass of $M = 0.7$\,M$_{\odot}$, 
thereby considering that our RR\,Lyrae stars can deviate 15\% from 
that value. 
This deviation propagates to a maximum deviation of 0.06 dex 
in a calculated $\log g(T,L,M)$. 

A few observational determinations of the absolute magnitude $M_V$ and 
mass $M$ of horizontal-branch and RR\,Lyrae stars exist, 
such as those of de Boer et\,al.{} (1995), Moehler et\,al.{} (1995), 
de Boer et\,al.{} (1997), 
Moehler et\,al.{} (1997), Tsujimoto et\,al.{} (1998), and Gratton (1998). 

We return to the effect of incorrect values 
of these reference parameters in Sect.\,\ref{errorbudget}.


\vspace*{1mm}

The run of $T_{\rm eff}$, $\log g_{\rm BJ}$, and $R$ derived as described 
above are shown for three of our stars in Fig.\,\ref{parcurve}. 
The shape of these curves is very similar to those presented by 
van Albada \& de Boer (1975). 
But the data of Fig.\,\ref{parcurve} is less noisy because 
the photometry with CCDs is faster and more precise than possible 
with the slower photomultipliers of that time, 
and because of the simultaneity of the {B\sc usca} measurements. 

\begin{table}
\caption[]{Cycle averages of several RR Lyr star parameters}
\begin{center}
\begin{tabular}{llccc}
\hline
\hline
Star \hspace*{3mm} & $\langle L \rangle$\hspace*{2mm} & $\langle T_{\rm eff}\rangle$ & $\log \langle g(T,L,M) \rangle$ & $\log \langle g_{\rm BJ} \rangle$ \\
\hline
\object{RR Gem} &  43.5 & 6647 & 2.84 & 3.33 \\
\object{TW Lyn} &  31.6 & 6196 & 2.85 & 3.61 \\
\object{AS Cnc} &  49.3 & 6592 & 2.77 & 3.46 \\
\object{SY Ari} &  50.0 & 6059 & 2.61 & 2.93 \\
\object{SZ Gem} &  52.9 & 6739 & 2.78 & 3.38 \\
\object{BH Aur} &  46.4 & 5997 & 2.63 & 3.04 \\
\object{TZ Aur} &  50.3 & 5934 & 2.58 & 2.86 \\
\hline
\object{AR Per}\tablefootmark{a} & 56.3 & 6030 & 2.55 & 3.66 \\
\object{CI And}\tablefootmark{b} & 47.7 & 6270 & 2.69 & 3.19 \\
\object{BR Tau}\tablefootmark{c} & 44.  & 5980 & 2.65 & 2.92 \\
\object{AA Cmi}\tablefootmark{c} & 45.  & 6340 & 2.73 & 3.28 \\
\hline
\end{tabular}
\end{center}

\tablefoot{
$L$ in L$_{\odot}$; $T$ in K; $g$ in cm\,s$^{-2}$. 
\ $L$ derived from $y$, $A_V$, $d$, and B.C.\\
\tablefoottext{a}{Light curve sparsely measured (see Fig.\,\ref{flightcurves}).}
\tablefoottext{b}{Light curve with a large gap in the descending branch; interpolated.}
\tablefoottext{c}{Relatively large observing gaps (see Fig.\,\ref{flightcurves}); interpolation uncertain.}
}
\label{ttabaver}
\end{table}

\subsection{Phase resampling in steps of 0.02}
\label{resampling}

To facilitate the calculation of averages over the phase 
and additional analyses, 
we resampled the curves of $L$, $T_{\rm eff}$, $R$, and $\log g_{\rm BJ}$ 
as derived from the observational data using steps of $\Delta\Phi=0.02$ 
(starting with $i=0$ near maximum light). 
These resampled values (given henceforth with subscript 0.02) 
can be easily integrated (fixed time step). 
The result of these resamplings can be seen as curves in 
Figs.\,\ref{tefflloop} and \ref{parcurve} (and in further figures). 

Time averages over the cycle of our stars have been calculated based on the 
0.02 phase stepped lightcurves. 
For a few stars, the lightcurve coverage was not complete and we made 
reasonable interpolations in the gaps to obtain the time averages. 

\subsection{Time-averaged quantities}
\label{timeaverage}

The structure of a ``non-pulsating RR Lyr star'' is given by the 
time-averaged parameters $\langle T_{\rm eff}\rangle$, 
$\langle L\rangle$ and $\langle g \rangle$. 
The time-averaged gravity, $\langle g \rangle $, 
has in a purely periodic system zero contribution from $d^2R/dt^2$. 

Oke and collaborators and van Albada \& de Boer felt that 
since $g$ varies strongly, the ``average'' value 
is best derived from the quiet part of the cycle, 
i.e., from $0.3<\Phi<0.85$. 
However, Fig.\,\ref{favvrad} makes clear that this has shortcomings. 
In that phase interval, $\log g(T,L,M)$ is (for most stars) 
lower than average by between 0.1 and 0.2 dex,  
because the atmosphere is then slowly but continuously contracting 
(but see Sect.\,\ref{secggeff}). 

The time-averaged quantities of essential parameters were calculated from 
the resampled curves in steps of $\Delta \Phi=0.02$ (see above), 
the integral was derived as the sum of 50 values. 
Thus 
\begin{equation}
\langle T_{\rm eff} \rangle = \int T_{\rm eff} \ dt = \bigl(\Sigma_{i=1}^{50}\ (\,(T_{\rm eff})_{0.02})\,_i\bigr)\ /\ 50 
\end{equation}
and 
\begin{equation}
\langle L \rangle = \bigl(\Sigma_{i=1}^{50}\ (L_{0.02})\,_i\bigr)\ /\ 50 \ \ \ ,
\end{equation}
individual values of $L_{0.02}$ being calculated as described 
in Sect.\,\ref{parpulsating}. 
The averages of the two gravity versions are 
\begin{equation}
\langle g(T,L,M) \rangle = \bigl(\Sigma_{i=1}^{50}\ (\,g(T,L,M)_{0.02})\,_i\bigr)\ /\ 50
\end{equation}
and
\begin{equation}
\langle g_{\rm BJ} \rangle = \bigl(\Sigma_{i=1}^{50}\ (\,(g_{\rm BJ})_{0.02})\,_i\bigr)\ /\ 50 \ \ \ .
\end{equation}

\begin{figure}
\resizebox{\hsize}{!}{\includegraphics{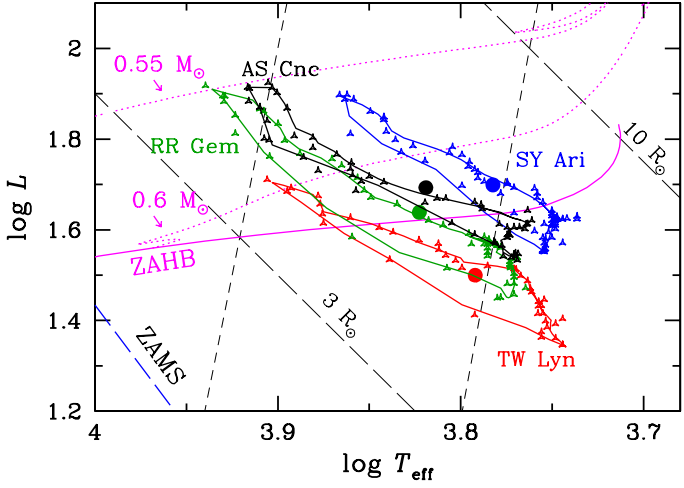}}
\caption[Loops in Teff and L]{For four RR\,Lyrae 
(\object{TW Lyn}, \object{RR Gem}, \object{AS Cnc}, and \object{SY Ari}), 
the run of $T_{\rm eff}$ and $L$ is shown in an HRD. 
The originally derived values are plotted as 3-pointed stars, 
the resampled curve (in steps of 0.02 in phase, Sect.\,\ref{resampling}) 
is shown as a full line. 
The mean values over the cycle (from Table\,\ref{ttabaver}) 
are given as~$\bullet$.
The background data (ZAMS, metal-poor ZAHB, 
the evolution tracks of a 0.55 and a 0.6 M$_{\odot}$\ HB star) 
are taken from de Boer \& Seggewiss (2008; Fig 10.9). 
Two lines of constant $R$ are indicated. 
The almost vertical dashed lines mark the approximate location 
of the pulsational instability strip from Gautschi \& Saio (1995). 
The stellar surface parameters of the RR\,Lyrae show hysteresis.
}
\label{tefflloop}
\end{figure}

Time-averaged values for 11 stars are given in Table\,\ref{ttabaver} 
and are for four stars included in Fig.\,\ref{tefflloop}. 
It can immediately be seen that the values of 
$\langle g(T,L,M) \rangle$ and $\langle g_{\rm BJ} \rangle$ 
are quite dissimilar. 
We discuss this further in Sect.\,\ref{secggeff}. 

\subsection{Error budget}
\label{errorbudget}

In our analyses, the following uncertainties had to be taken into account.

Photometric errors are, given the instrumental characteristics of {B\sc usca}, 
small. 
We estimated the uncertainties in $y$ to be 0.03 mag. 
The bolometric corrections are small and introduce errors of up to 2\%.
This leads to a total typical {\sl photometric} uncertainty in $L$ of 3\%. 
In the colour indices $b-y$, $m_1$, and $c_1$, the uncertainties are 0.02 mag 
(smaller than the error in $y$ 
because of the simultaneity of the measurements).

We used distances from the literature. 
The effect of their uncertainties on $L$ is given below. 

To obtain $T_{\rm eff}$ and $\log g_{\rm BJ}$, 
the indices $b-y$ and $c_1$ were used with the grid of Clem et\,al. (2004). 
The grid is not perfect and reading $T_{\rm eff}$ and $\log g_{\rm BJ}$ from 
that grid by eye, we assessed that 
(based on readings performed three times for the entire cycle for one star) 
the uncertainty in $T_{\rm eff}$ is about 1\%,
and in $\log g_{\rm BJ}$ about 0.05 dex. 
Since the effects of metallicity, [M/H], 
are only significant for $T_{\rm eff} < 6000$\,K, 
which occurs for only 3 of our less well-observed stars near minimum light, 
we are confident that our error estimates are only at those temperatures 
affected by a possibly small mismatch in [M/H]. 

Given the measurement uncertainties in $L$ (3\%) and $T_{\rm eff}$ (1\%), 
it follows that the uncertainty in the calculated $R$ values is 4\%. 
If, in addition, a distance were incorrect by +10\% that would give 
an additional offset for a given star in $R$ of +5\%. 

As mentioned in Sect.\,\ref{parpulsating}, 
we assumed a fixed mass for all stars of 0.7\,M$_{\odot}$, 
and indicated a margin of error of 15\%. 
This error would lead for a given star to an offset of 0.06 dex 
in a calculated $\log g(T,L,M)$. 

The values of $\log g(T,L,M)$\ calculated from Eq.\,\ref{elogMTgL} 
have a measurement error of about 8\% or 0.03 dex. 
A distance error of +10\% leads to an offset in $\log g(T,L,M)$\ of $-$0.04, 
an error in the mass $M$ of +10\% to an offset in $\log g(T,L,M)$\ of +0.04. 

Finally we note that the distances of RR\,Lyrae have essentially all 
been derived from a reference $M_V$ (distance from distance modulus), 
in some cases adjusted for metal content, [Fe/H]. 
For a review of the latest calibrations we refer to Sandage \& Tamman (2006). 
However, in spite of all calibration efforts, 
$M_V$ has a spread in $M_V$ of between 0.2 and 0.4 mag for a given [Fe/H] 
or an uncertainty of even up to 1.0 mag if [Fe/H] is unknown.  
Moreover, given the evolution of RR\,Lyrae stars on the HB, 
a single and universal value of $M_V$ does not exist anyway; 
the evolution of an HB star may lead to an increase in $\log L$ of 
0.5 dex when reaching the terminal-age HB 
(see, e.g., de Boer \& Seggewiss 2008; their figure~10.9). 
Therefore, if a considered star were brighter than the chosen reference value, 
its derived distance modulus and thus distance 
would be too small. 
If the true $M_V$ of the star is unbeknown 
to be 0.25 mag brighter than the reference value, 
the distance (as given in Table\,\ref{tabobs}) 
will turn out to be too small by 12\%. 
This affects the derived $L$, then being too small by 0.10 dex. 

\begin{figure}
\resizebox{\hsize}{!}{\includegraphics{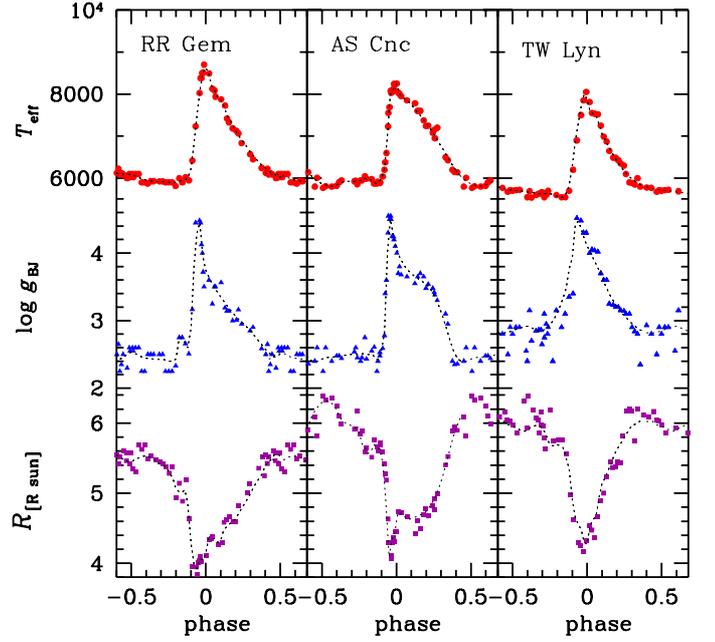}}
\caption[parameter curves]{Run of the parameters $T_{\rm eff}$, 
$\log g_{\rm BJ}$, and $R$ (derived as described in Sect.\,\ref{parpulsating}) 
during the pulsational cycle of \object{RR Gem}, \object{AS Cnc}, 
and \object{TW~Lyn} using the calibration grid by Clem et\,al. (2004). 
The  thin dashed lines are the values resampled at $\Delta\Phi$ = 0.02 
(see Sect.\,\ref{resampling}). 
}
\label{parcurve}
\end{figure}

The mean values of Table\,\ref{ttabaver} 
were derived from a large number of individual measurements, 
so the observational scatter is basically cancelled. 
What remains is the error related to the calibration grid and 
also the error in the assumed distance and/or $M_V$ 
(affecting $\log g_{\rm BJ}$ and $L$ as just indicated, as well as $R$). 

\section{Hysteresis of the parameters $T_{\rm eff}$, $R$, and $\log g_{\rm BJ}$}

The values of the atmospheric parameters for all points in the 
cycle were derived from the photometry as described above. 
In Fig.\,\ref{tefflloop} we show the run of $T_{\rm eff}$ and $L$ 
of four of our stars in an HRD. 
These parameters exhibit loops through the diagram 
extending over and beyond the instability strip. 
The loopings are reminiscent of hysteresis. 
In the older literature, the lagging behind in the change of some parameter 
compared to another parameter is called ``phase-lag'' (see, e.g., Cox 1980). 

The behaviour of $T_{\rm eff}$ and $\log g_{\rm BJ}$ against phase 
(Fig.\,\ref{parcurve}) shows remarkable features. 
For all stars, $\log g_{\rm BJ}$ begins to increase 
about $\Delta\Phi=0.05$ before $T_{\rm eff}$ starts to increase.
Furthermore, 
$\log g_{\rm BJ}$ reaches its highest value 
when $T_{\rm eff}$ has increased about halfway to its maximum value. 
And when $T_{\rm eff}$ reaches its maximum, 
$\log g_{\rm BJ}$ is already decreasing (at phase $\simeq 0$). 
The resulting value of $R$ 
(that of Fig.\,\ref{parcurve} is derived from Eq.\,\ref{elogLRT}) 
indicates that $R$ is smallest when $\log g_{\rm BJ}$ is largest. 

The run of $R$ shows that RR\,Lyrae are relatively extended stars 
for about half of the cycle ($0.25 < \Phi < 0.75$), 
while large changes in radius take place between $0.75 < \Phi < 1.25$, 
with the most rapid change occurring between $0.9 < \Phi < 1.15$. 
At $\Phi \simeq 0.9$, a drastic reduction in the derived $R$ takes place 
(the atmosphere apparently starts to collapse), 
reaching the smallest $R$ near $\Phi\simeq 0.95$. 
At smallest $R$ the atmosphere must have a high density. 
Up to the highest density, $T_{\rm eff}$ is at the same level 
as in the quiet phase leading up to the collapse. 
Soon after $\Phi\simeq 0.95$, $R$ increases again. 

In general, 
when $y$ is brightest $T_{\rm eff}$ is at its highest and $R$ is very small; 
but immediately after minimum light, at the high peak in $\log g_{\rm BJ}$, 
$R$ is at its smallest. 

\section{Spectral line strengths}
\label{speclines}

For three of our stars (\object{TZ Aur}, \object{RR Gem}, \object{RS Boo}), 
spectra were obtained 
(in January 2006, independently of the Str\"omgren-photometry) 
with the focal reducer spectrograph at the 1\,m telescope of the 
Observatory ``Hoher List'' of the AIfA. 
Radial velocities could not be obtained with this instrument. 
The spectra have a resolution of $\sim$3~\AA. 
Because of telescope size and focal-reducer set-up, 
exposure times had to be rather long, 
of the order of 3\% of the period, leading to phase smearing. 
We determined the equivalent withs of the lines of H$\alpha$, H$\beta$, 
H$\gamma$, Na\,{\sc i}\,D, and Ca\,{\sc ii}\,K. 

For RR\,Gem and TZ\,Aur, we show (Fig.\,\ref{rrgemspec}) 
how the strengths of the H$\gamma$ and Ca\,K lines vary during the cycle. 
Results for RS\,Boo are not shown because the lightcurve coverage was poor. 

\noindent
{\bf H}$\gamma$. 
The strength of Balmer lines is governed by $T_{\rm eff}$ as expected 
since the Balmer level can be populated 
only during the higher $T_{\rm eff}$ part of the RR\,Lyrae cycle. 
The temporal behaviour of $W$(H$\gamma$) and $T_{\rm eff}$ 
(Fig.\,\ref{rrgemspec}) is indeed very similar. 

\noindent
{\bf Ca\,{\sc ii}}\,K. 
The strength of Ca\,K is predominantly set by the gas density, i.e., 
the ionization balance forces Ca\,{\sc ii} to be reduced at higher density 
(when $R$ is smallest). 
The curves of $W$(CaK) and $R$ are very similar in terms of 
their temporal behaviour (clearly for RR\,Gem). 

The Na\,{\sc i}\,D absorption is very weak and provides little information, 
except that Na\,{\sc i}\,D is strong when $T_{\rm eff}$ is at its lowest; 
the increased strength is caused by the shift of the ionization balance 
towards the neutral state at lower $T_{\rm eff}$. 

\section{The variation in the pulsation velocity $V_{\rm pul}$} 

\subsection{The coarse variation in $V_{\rm pul}$}
\label{varvrad}

Figure\,\ref{parcurve} shows for three stars the change in radius, $R$, 
as derived from Eq.\,\ref{elogLRT}. 
One can thus calculate the velocity of the stellar atmosphere 
(the ``pulsation velocity'', $V_{\rm pul}$). 
We used 
\begin{equation}
 V_{\rm pul} = \frac{\Delta R}{\Delta t} = 
  \frac{R_{\Phi}-R_{\Phi+\Delta\Phi}}{t_{\Phi}-t_{\Phi+\Delta\Phi}} \ \ \ ,
\label{evrad}
\end{equation}
where $R$ is at intervals of fixed value 
$\Delta \Phi= 0.02$ (see Sect.\,\ref{timeaverage}). 
To avoid being affected by peaks in the noise we applied 
a running triangular (1,2,1) smoothing to the curves of $R$ 
before calculating $V_{\rm pul}$. 
In Fig.\,\ref{favvrad}, $V_{\rm pul}$ (as well as the smoothed $R$) is given 
for eight stars with the best data. 
Note that $V_{\rm pul}$ plotted is that seen from the centre of the star 
(and is not a heliocentric value). 

\begin{figure}
\resizebox{\hsize}{!}{
\includegraphics{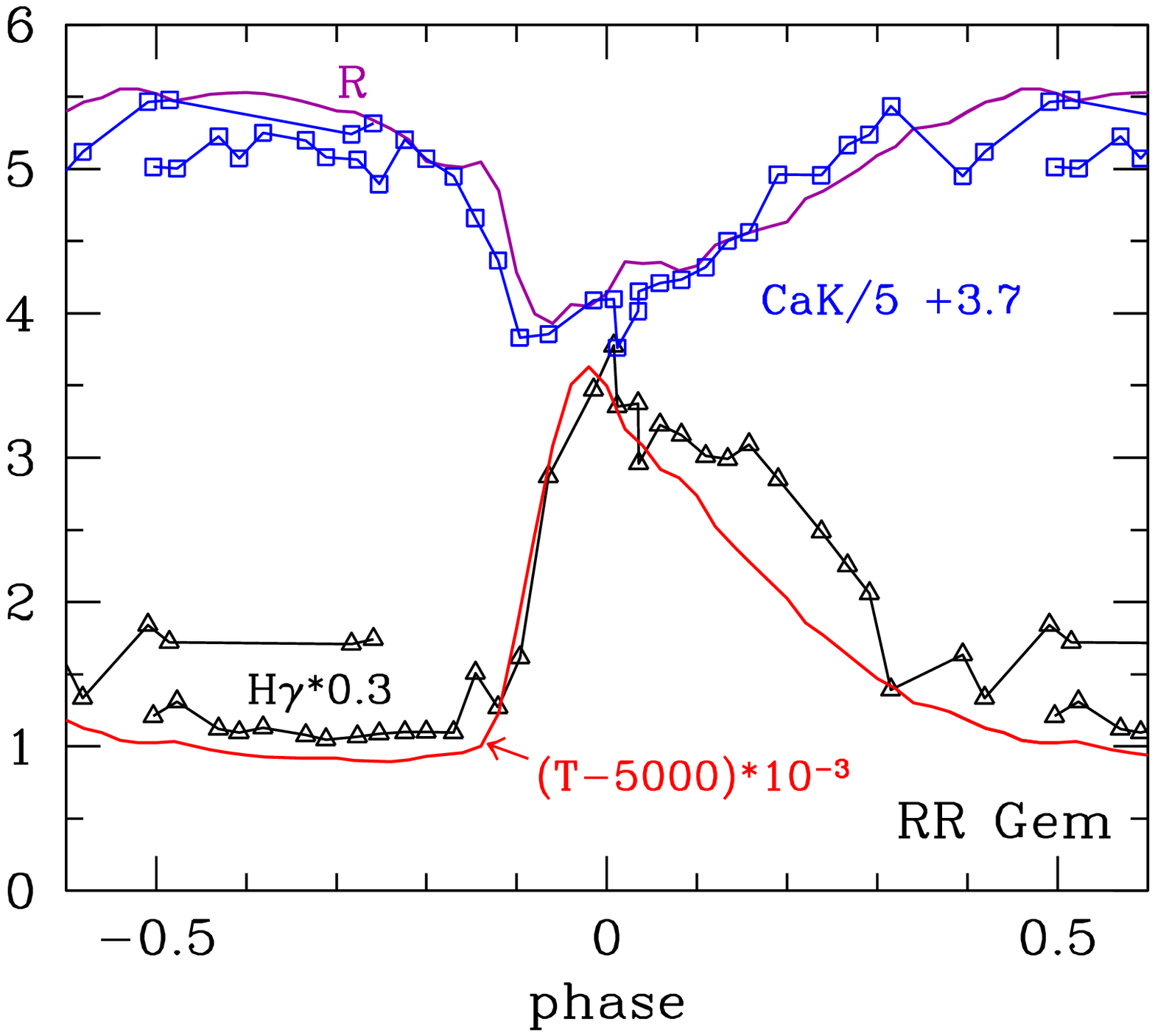} 
\includegraphics{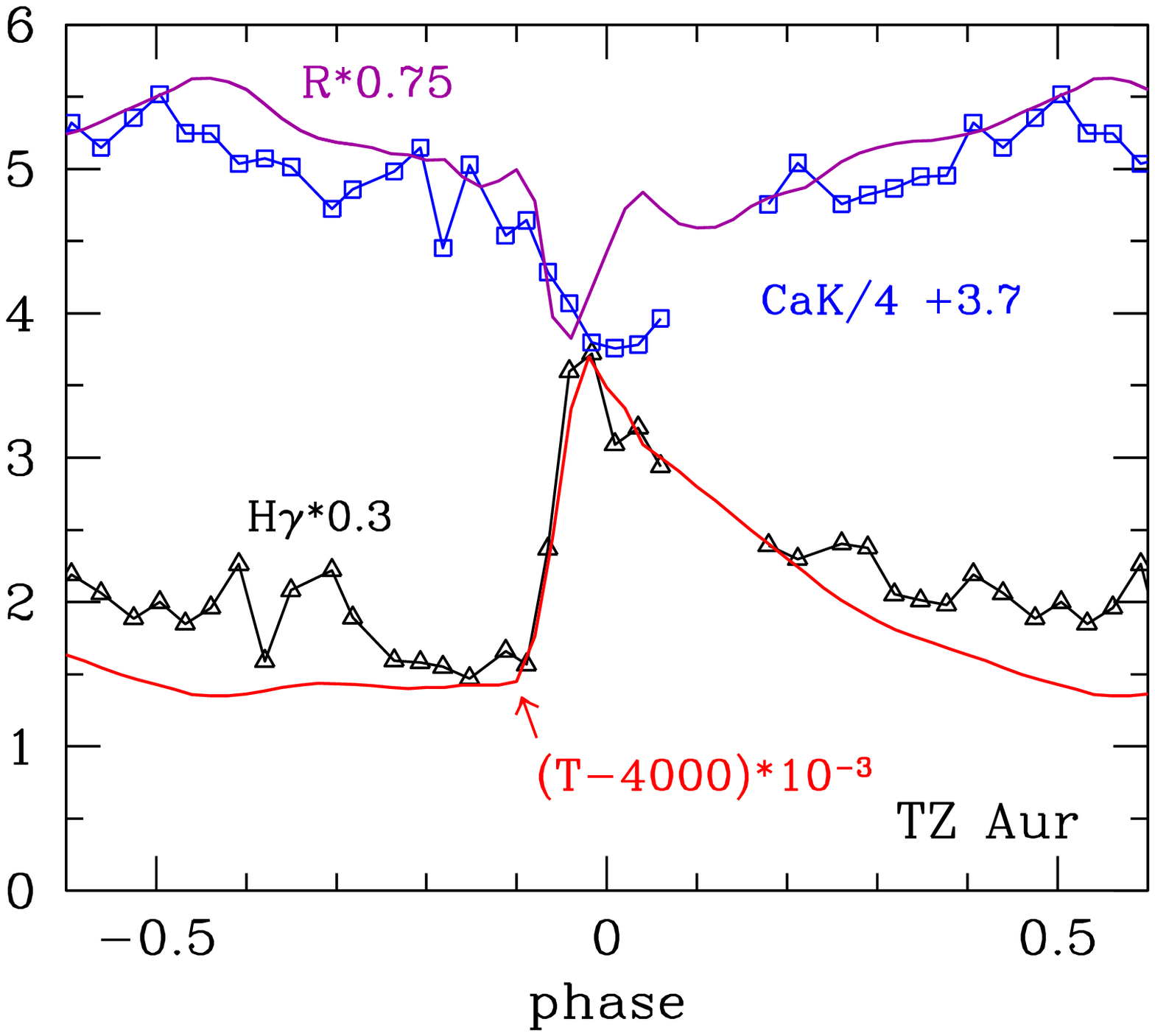}
}
\caption[spectral line strengths]
{For \object{RR Gem} and \object{TZ Aur}, the run of the equivalent widths $W$ 
of the spectral lines Ca\,{\sc ii}\,K and H$\gamma$ are shown and compared 
with the parameters $R$ and $T_{\rm eff}$ (as in Fig.\,\ref{parcurve}). 
The vertical scale is that of $R$ [R$_{\odot}$] of RR\,Gem. 
The values of the other parameters ($W$ in \AA\ and $T_{\rm eff}$ in K) have 
been renormalized to fit this vertical scale (see the labels to the curves). 
For RR\,Gem, H$\gamma$ is proportional to $T_{\rm eff}$ and 
is strongest when $T_{\rm eff}$ is largest (stronger excitation); 
Ca\,K is weakest when the radius $R$ is smallest 
(at constant $T$ and higher density, $n_e$ shifts the ionization balance). 
The spectral observations of RR\,Gem come from two nights 
in the same time frame as but independent of the Str\"omgren photometry 
(explaining the overlap of data in phase). 
In TZ\,Aur the relation of $R$ with Ca\,K is less clear.
The spectra of TZ\,Aur did not cover the entire light curve 
(gap near $\Phi = 0.15$).
}
\label{rrgemspec}
\end{figure}

\begin{figure*}
\resizebox{\hsize}{!}
{\includegraphics{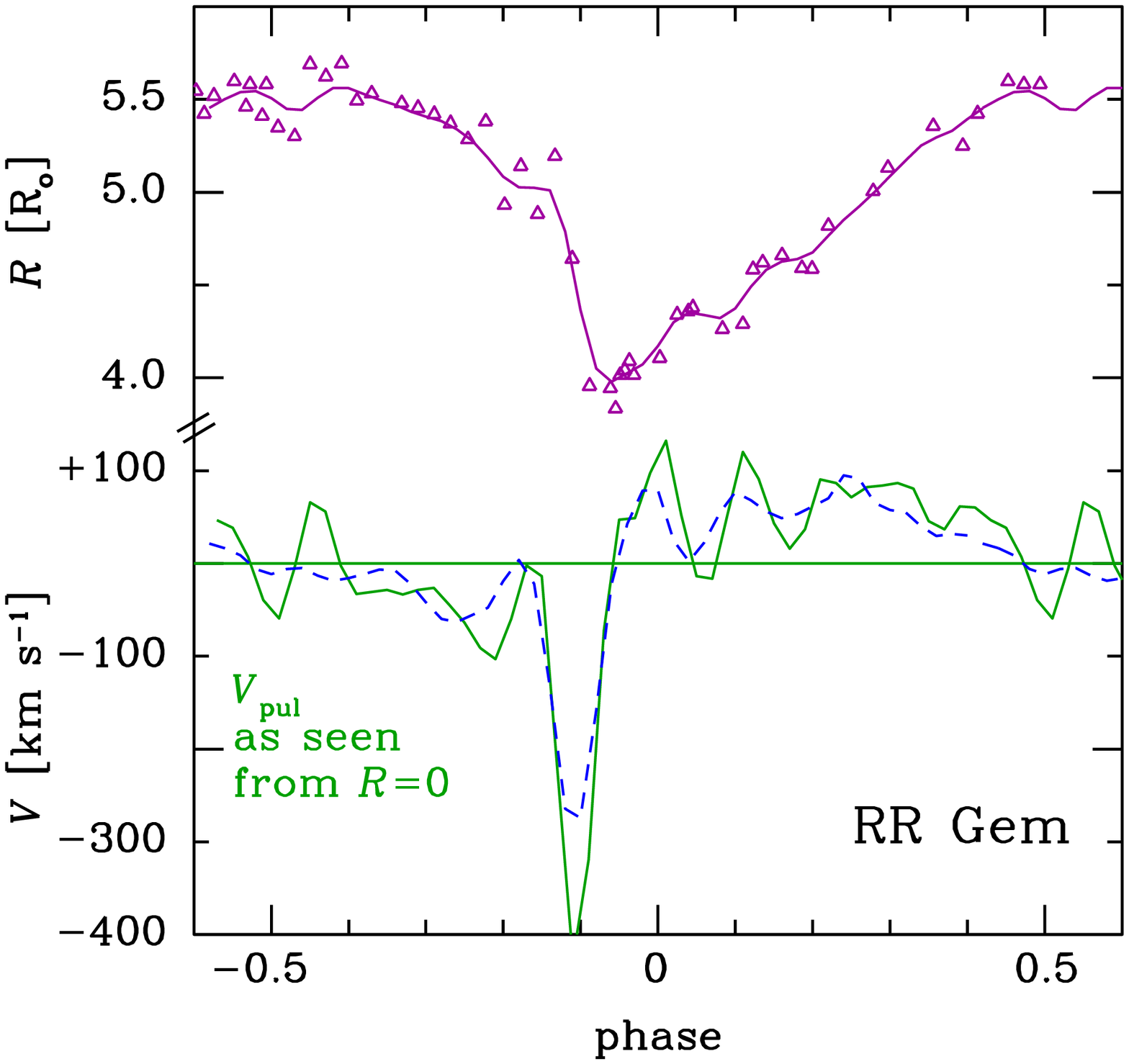}
\includegraphics{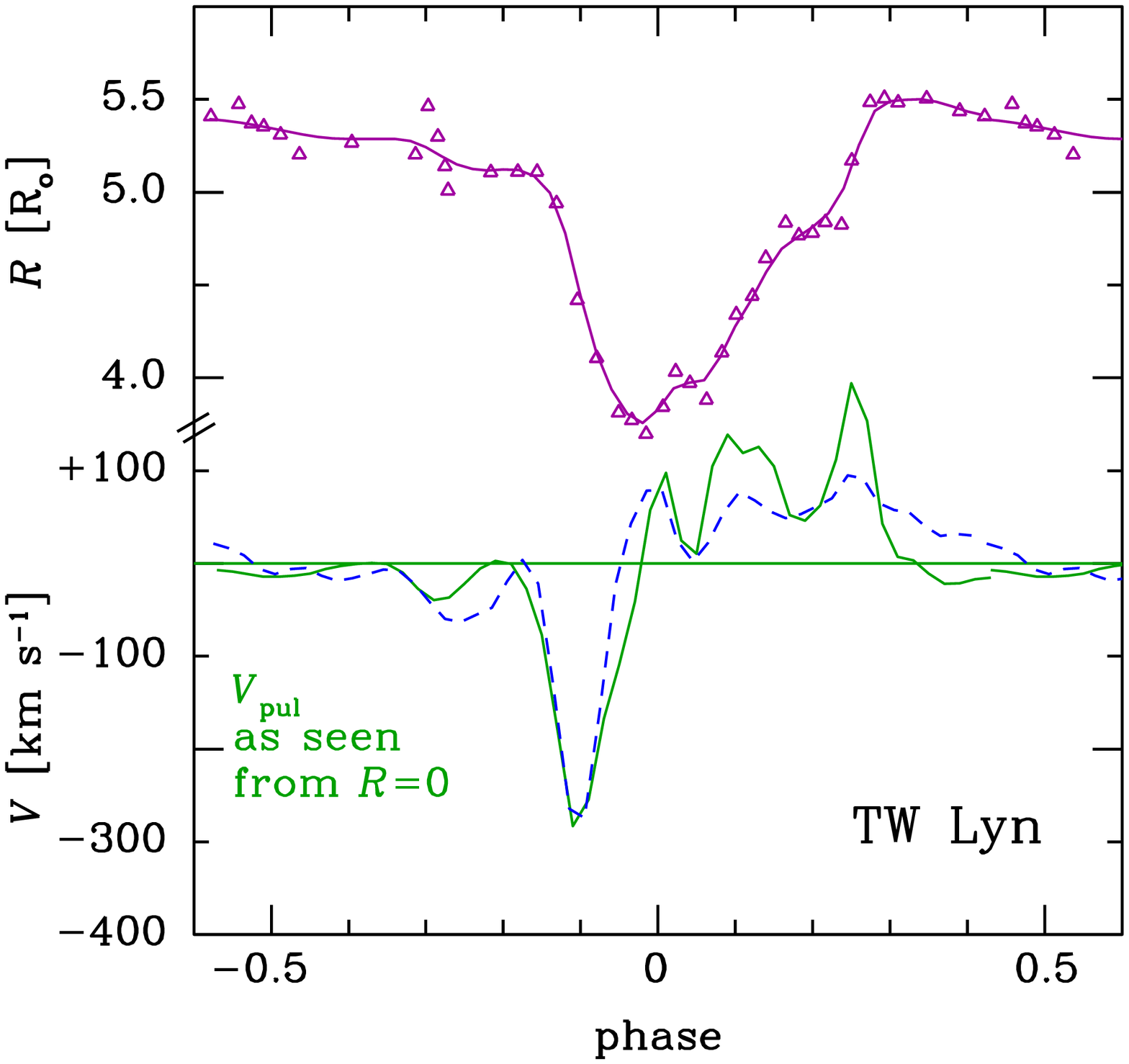}
\includegraphics{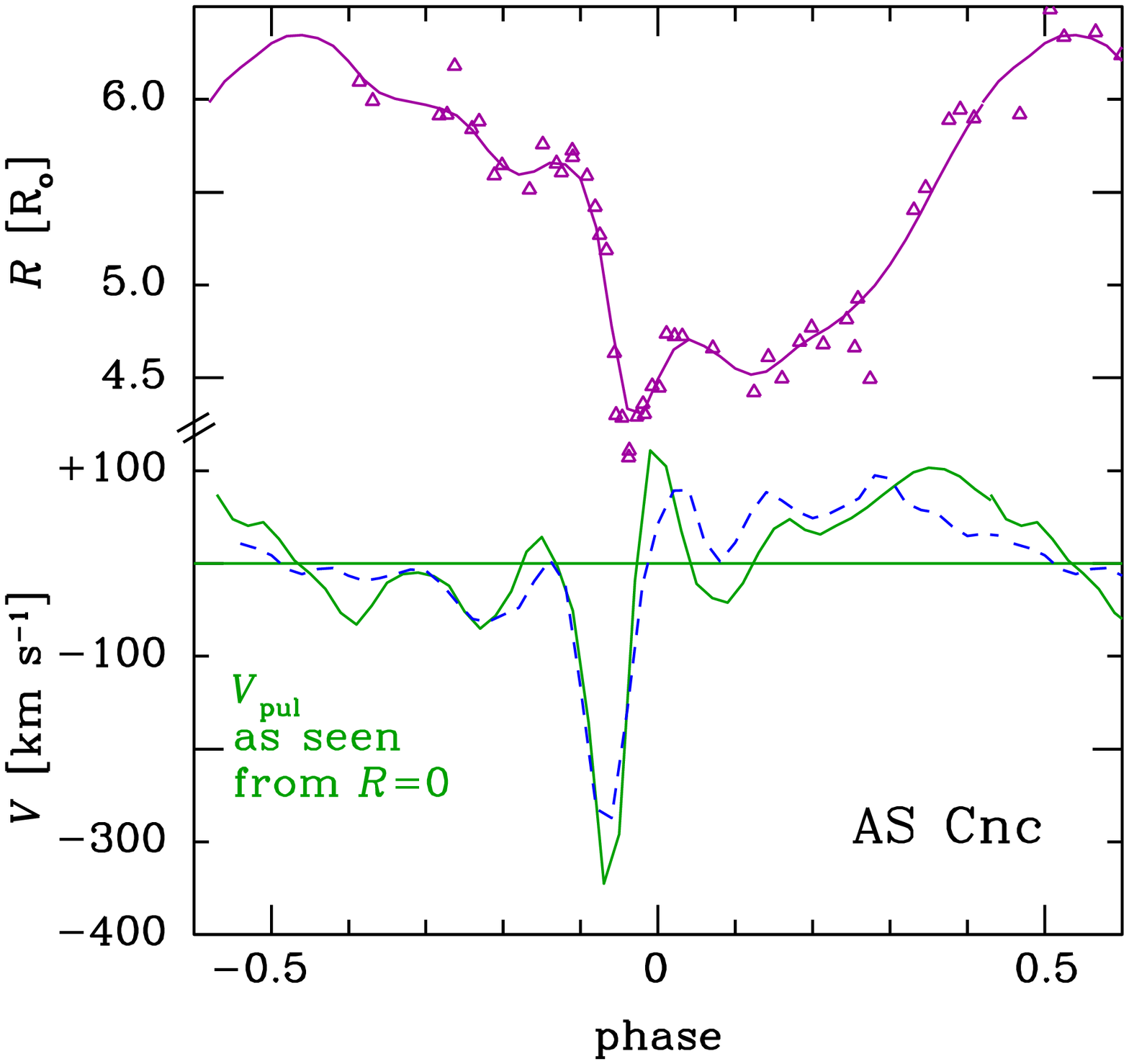}
\includegraphics{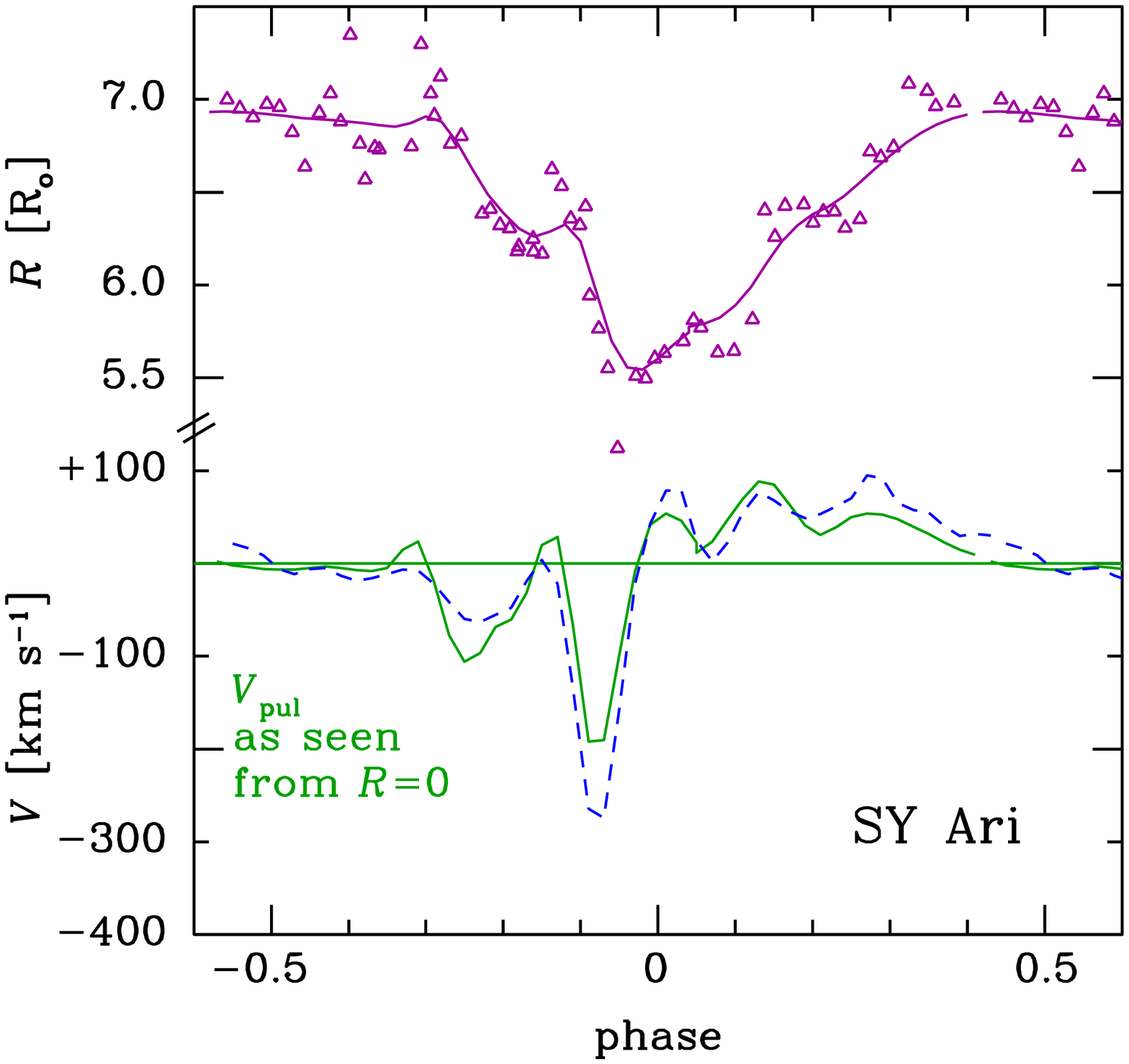}}
\resizebox{\hsize}{!}
{\includegraphics{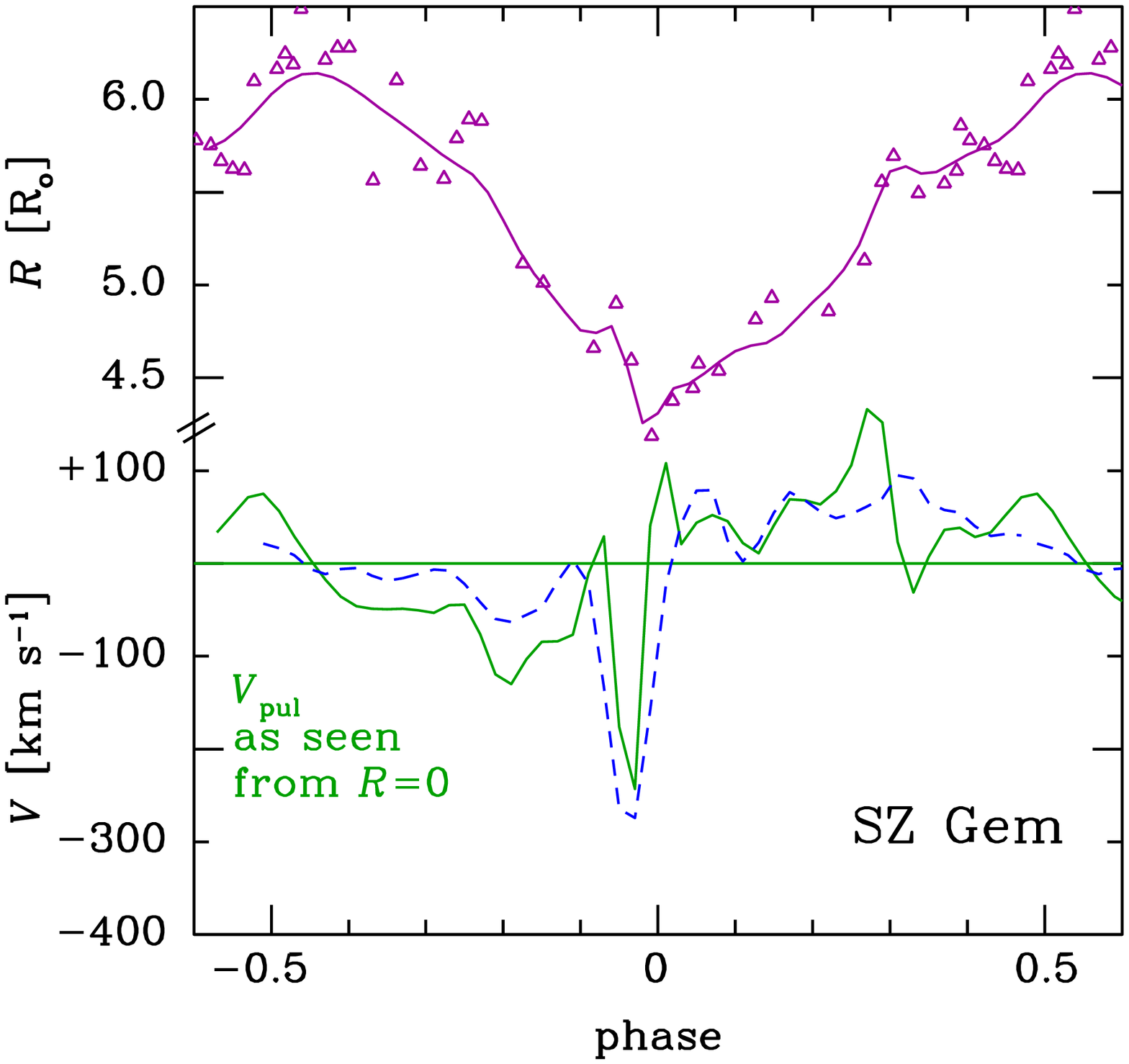}
\includegraphics{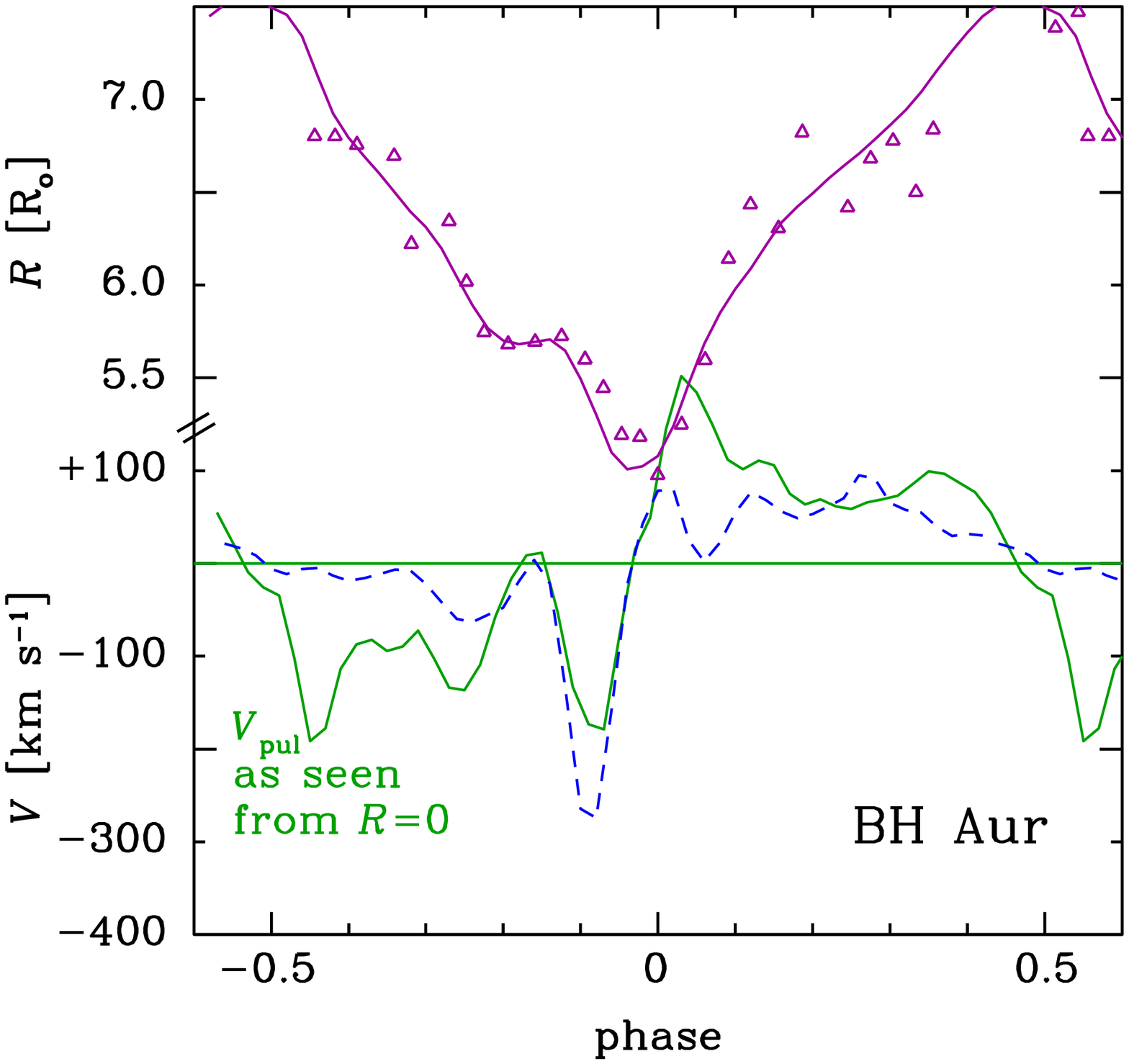}
\includegraphics{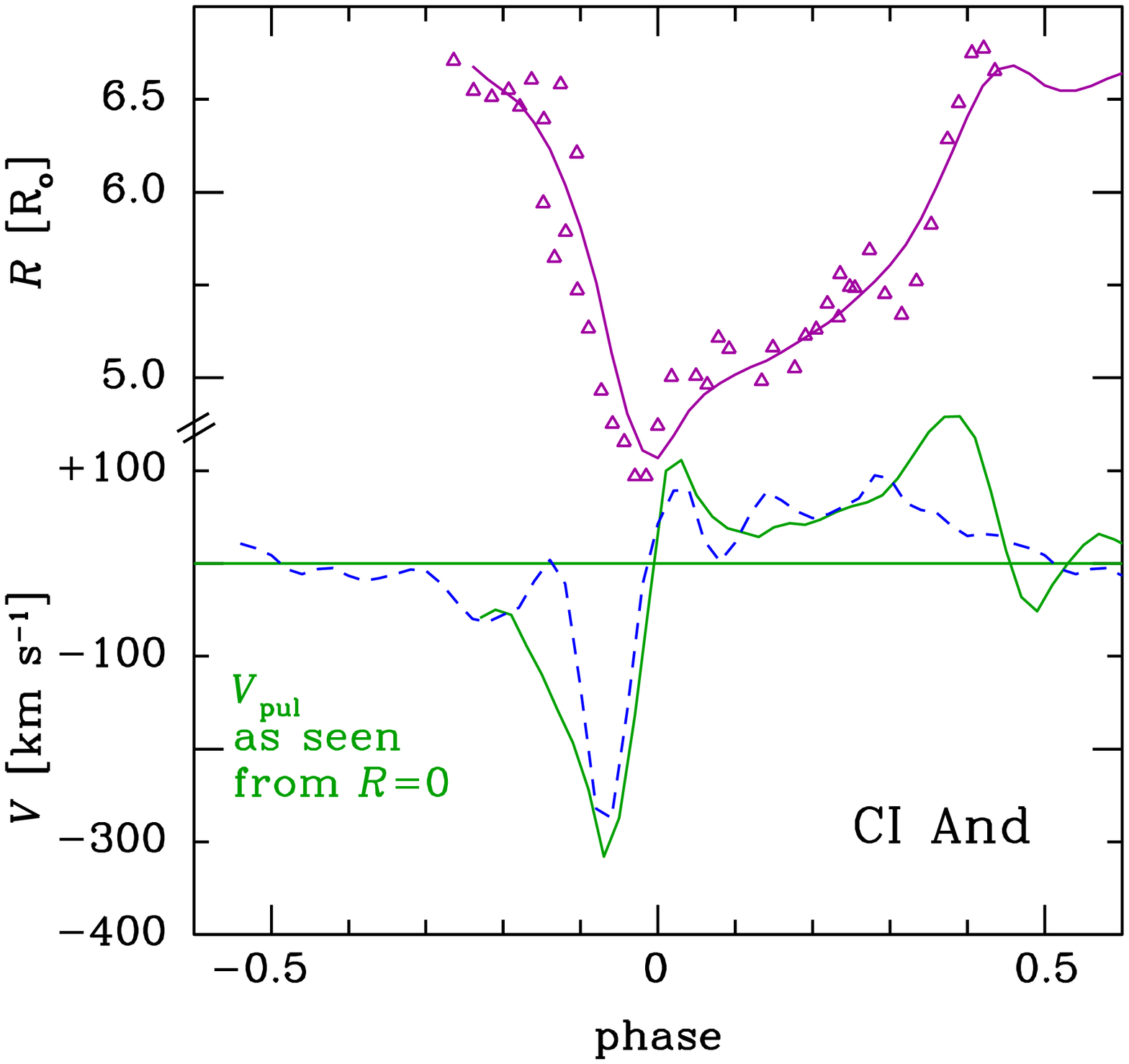}
\includegraphics{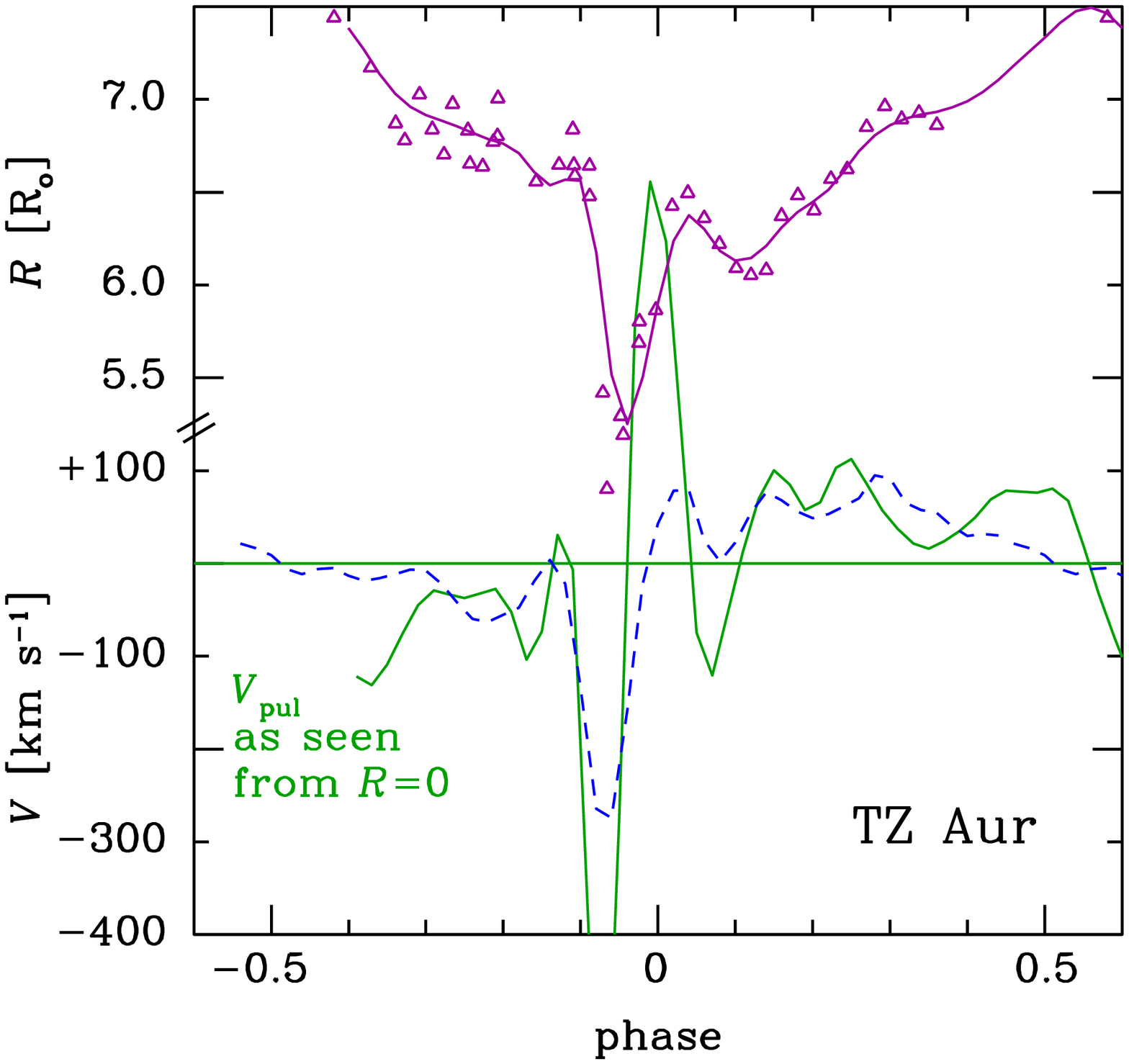}}
\caption[curves with average vrad]
{The change in $V_{\rm pul}$ is shown together with the change in $R$ 
(note that the scale of $R$ differs from panel to panel). 
{\sl Top row}: the ``well behaved'' and well-observed stars. 
{\sl Bottom row}: the stars with less well behaved $V_{\rm pul}$-curves 
(partly due to poor light-curve coverage and to interpolation). 
{\bf Top curve and data in each panel}: 
The values of $R$ were calculated from photometric data 
using Eqs.\,\ref{elogLRT} and \ref{elogMRg} (as in Fig.\,\ref{parcurve}). 
The resampled data ($\Delta\Phi=0.02$) 
were used to derive the variation in $R$ over these intervals. 
The curves shown give the resampled values 
after an additional running triangular (1,2,1) smoothing. 
Note that, for several stars, 
the period was not fully covered by our observations. 
{\bf Bottom curves in each panel}: 
The run of $V_{\rm pul}$ as derived from the smoothed $R$-curve is shown.
To guide the eye, a line at $V_{\rm pul}$=0 km\,s$^{-1}$ is added. 
The average of the $V_{\rm pul}$ curves of the 
four well observed stars (the stars shown in the top row) 
has also here been added as a dashed line 
(for each star a small appropriate phase shift was added). 
Clearly, the wiggles in $V_{\rm pul}$ are not noise but are systematic 
$V_{\rm pul}$-variations. 
This is confirmed by the stars of the bottom row, 
whose $V_{\rm pul}$ curves were {\sl not} included in the calculated average. 
The average $V_{\rm pul}$ curve shows a rhythm with a period of about $P/7$. 
\label{favvrad}
}
\end{figure*}

The calculated variation in $V_{\rm pul}$ over the cycle 
(see Fig.\,\ref{parcurve}) has a general pattern. 
There is a pronounced minimum close to $\Phi\simeq 0.95$ 
where $V_{\rm pul} \simeq -150$ to $-400$ km\,s$^{-1}$, followed by 
a steep rise to a level of $V_{\rm pul} \simeq +50$ to 100 km\,s$^{-1}$. 
In the course of the cycle, $V_{\rm pul}$ then slowly decreases and slowly 
becomes negative reaching a level of $V_{\rm pul} \simeq -50$ km\,s$^{-1}$. 
A sudden additional decrease sets then in to reach the most negative value 
of $V_{\rm pul}$, the value this description started with. 
It shows that the atmosphere really collapses between $\Phi=0.9$ and $1.0$. 
Thereafter, the expansion is drastic 
but beyond $\Phi \simeq 0.05$ it is far more gradual 
(at $V_{\rm pul} \simeq 50$ km\,s$^{-1}$) as it decelerates. 

The behaviour of $V_{\rm pul}$ is similar to that of a model 
atmosphere as presented in the gedankenexperiment by Renzini et\,al. (1992). 
That gedankenexperiment had a star with a core and an envelope, 
in which the luminosity offered to the base of the envelope, 
$L_{\rm B}$, was manipulated. 
A gentle decrease in this $L_{\rm B}$ from a high $L_{\rm B}$ state 
could be accommodated by the envelope 
until some critical value of $L_{\rm B}$ at which the envelope collapsed. 
Then, in the gedankenexperiment, 
$L_{\rm B}$ was raised again and the now compact atmosphere 
slowly expanded until some other critical value of $L_{\rm B}$ was reached, 
now leading to a run-away expansion. 

In the case of RR\,Lyrae, the run of $L$ is controlled by the opacity 
in the He$^+$-He$^{++}$ ionization boundary layer. 
With increasing transmission of energy through the envelope, $L$ increases 
and the star expands until at some limit the envelope has become so tenuous 
that (since $L$ can no longer increase) it must collapse. 

Having determined the run of $V_{\rm pul}$ one can derive $\frac{d^2R}{dt^2}$. 
These acceleration values change considerably during the pulsational cycle. 
However, because of the uncertainty in all $V_{\rm pul}$ values 
due to opacity effects to be discussed in Sect.\,\ref{companddepth}, 
we refrain from commenting on these acceleration values. 

\subsection{The detailed variation in $V_{\rm pul}$ and a correlation with $P$}
\label{avvrad}

The small-scale variations in the curves of $V_{\rm pul}$ 
(Fig.\,\ref{favvrad}) might be regarded as due to photometric noise. 
However, on closer inspection there is significant structure. 

We selected the well-behaved $V_{\rm pul}$-curves 
to determine an average $V_{\rm pul}$-curve for an RR\,Lyr star. 
For that, we chose the stars RR\,Gem, TW\,Lyn, AS\,Cnc, and SY\,Ari 
(the top row panels in Fig.\,\ref{favvrad}). 
Before the curves can be added for averaging, small phase shifts were applied 
because the chosen $y_{\rm max}$ was not perfectly aligned with the 
epoch relation and the resampling at 0.02 intervals in phase 
may also have introduced small phase shifts.
Figure\,\ref{favvrad} shows the results for these stars. 
It is evident that the wiggles in the $V_{\rm pul}$-curves 
are very similar for these four stars. 
For the four less well-observed stars not included in the averaging, 
a comparison with the average $V_{\rm pul}$-curve shows 
that these stars have also the same behaviour in $V_{\rm pul}$. 

For all stars, the wiggles in the average $V_{\rm pul}$ indicate 
that after the collapse the envelope oscillates 
(continues to oscillate) with a rhythm of about 1/7 of the period $P$. 
This rhythm at $P/7$ is the same for all our stars, 
independent of their actual period. 
It is as if the envelope has a residual oscillation triggered by the collapse. 
The collapse itself occurs from $V_{\rm pul}$$\simeq 0$ through the highest 
collapse velocity back to $V_{\rm pul}$$\simeq 0$ km\,s$^{-1}$, 
which is a time equal to about $P/7$. 

One might suspect that $P/7$ is due to our resampling at 0.02 phase intervals, 
the period being (almost) exactly 7 of our data points. 
However, 
we note that Papar\'o et\,al. (2009) reported a specific role of the 
7th harmonic in one well-studied {\sc CoRoT} RR\,Lyr star. 
The {\sc CoRoT} data do not have the 0.02 phase resampling rhythm. 
Barcza (2003) found in the RR\,Lyr star SU\,Dra from photometry 
an ``undulation''' of $P/5$. 
It is unclear what the cause of these periodicities is. 

\section{Comparison with literature data and optical depth effects}
\label{companddepth}

We derived the time-dependent radius of each star from our photometry 
and from its change the velocity of the atmosphere. 
The analysis above was performed without considering 
analyses of other RR\,Lyrae data presented in the literature. 

A well-known method for deriving the radius of RR\,Lyrae stars 
is the Baade-Wesselink method, 
where the radial velocities observed over the cycle are integrated 
to infer the variation in radius (see, e.g., Liu \& Janes 1990). 
This method has been applied to several RR\,Lyrae stars. 

\subsection{Comparison with other results}

For two of our stars (\object{RR Gem} and \object{AR Per}), cycle velocities 
were presented by Liu \& Janes (1989). 
Their velocities were obtained by cross-correlating their RR\,Lyrae star 
spectra with spectra of two standard stars, 
one of spectral type F6\,IV and one of A2\,IV. 
The values they derived from the two standard stars are very similar so, that 
their final velocity curves are based on just the standard of type A2\,IV. 

Our velocity curve for RR\,Gem has a shape quite different from the one 
of Liu \& Janes (1989). 
Note that in their figure\,5 they give the {\sl observed radial} velocity, 
whereas in our Fig.\,\ref{favvrad} 
the velocity as seen from the stellar centre is given. 
Their velocity curve shows a large speed away from the stellar centre 
at $\Phi\simeq 0$ 
followed by a gradual slowing down to approach the minimum in velocity 
at $\Phi\simeq 0.7$. 
The velocity curves of other RR\,Lyrae stars of Liu \& Janes 
show the same behaviour.  
The amplitude in their velocity curves is $\simeq$60 km\,s$^{-1}$. 
Adopting the usual geometry factor $p=1.32$ 
(see Liu \& Janes 1990; but see Fernley 1994) 
to convert from observed radial velocity to pulsational velocity, 
their velocities translate 
into an amplitude in $V_{\rm pul}$ of $\simeq$80 km\,s$^{-1}$. 

On the other hand, our stars have amongst themselves similar behaviour. 
Their outer layers experience during the cycle 
only a modest change in velocity 
except for a rapid fall toward the stellar centre as of $\Phi \simeq 0.85$ 
to a fastest centre approach at $\Phi \simeq 0.95$, 
followed by a rapid decline in velocity 
to reach a modest expansion velocity near $\Phi \simeq 0.05$. 
In our data, the full amplitude of the velocity $V_{\rm pul}$ 
is $\simeq$300 km\,s$^{-1}$. 

\begin{figure*}
\resizebox{\hsize}{!}{
\includegraphics{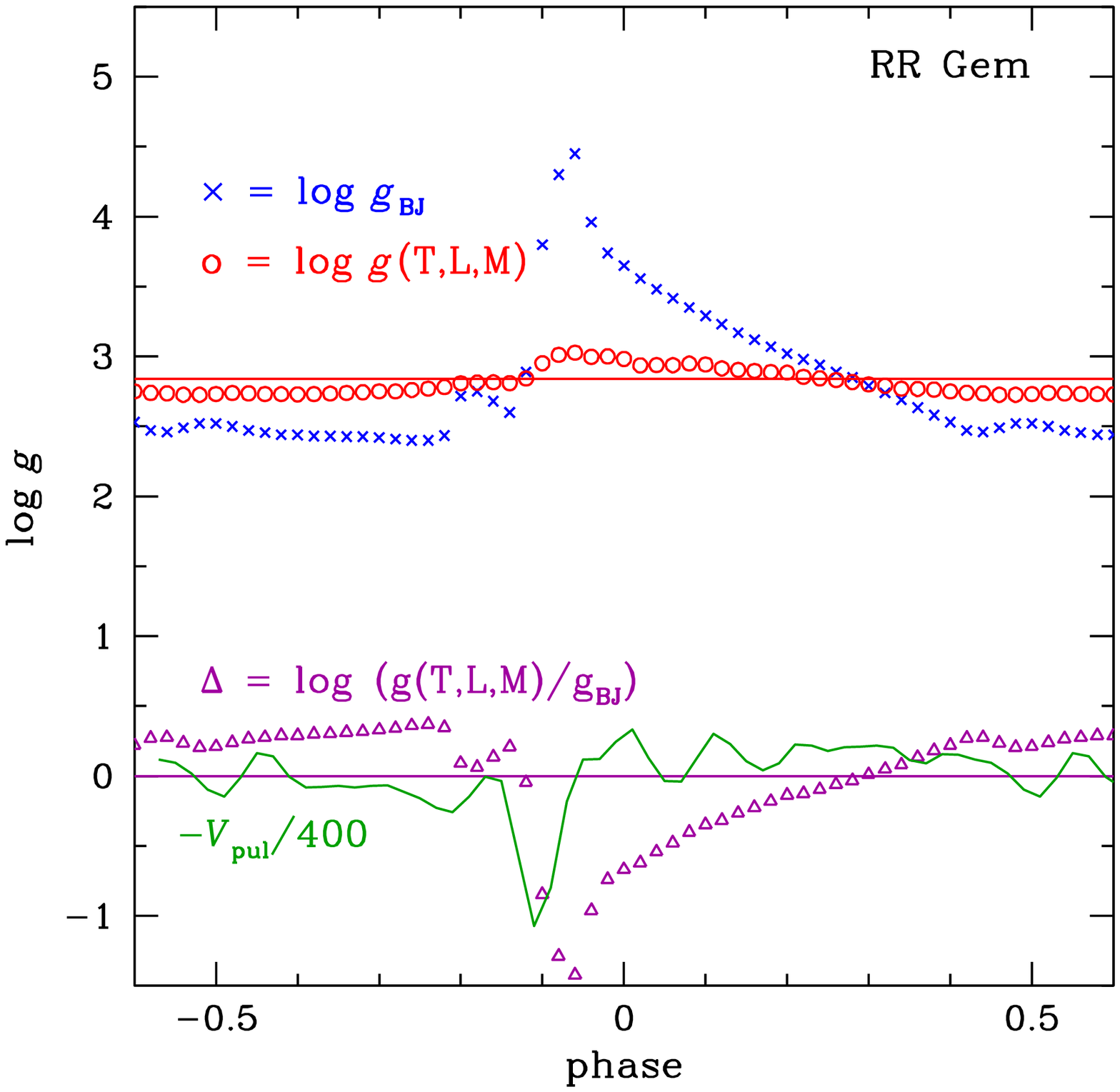}
\includegraphics{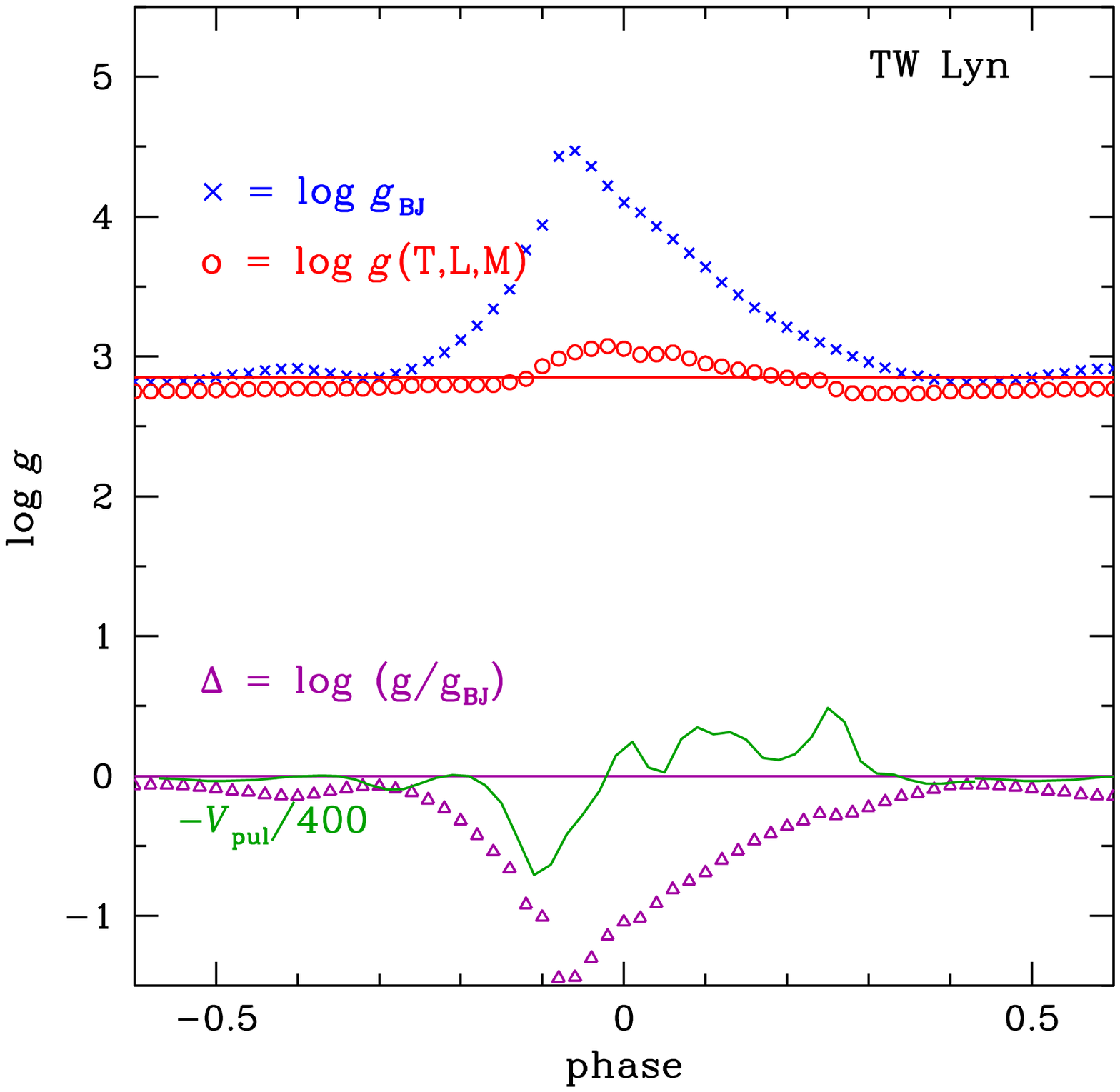}
\includegraphics{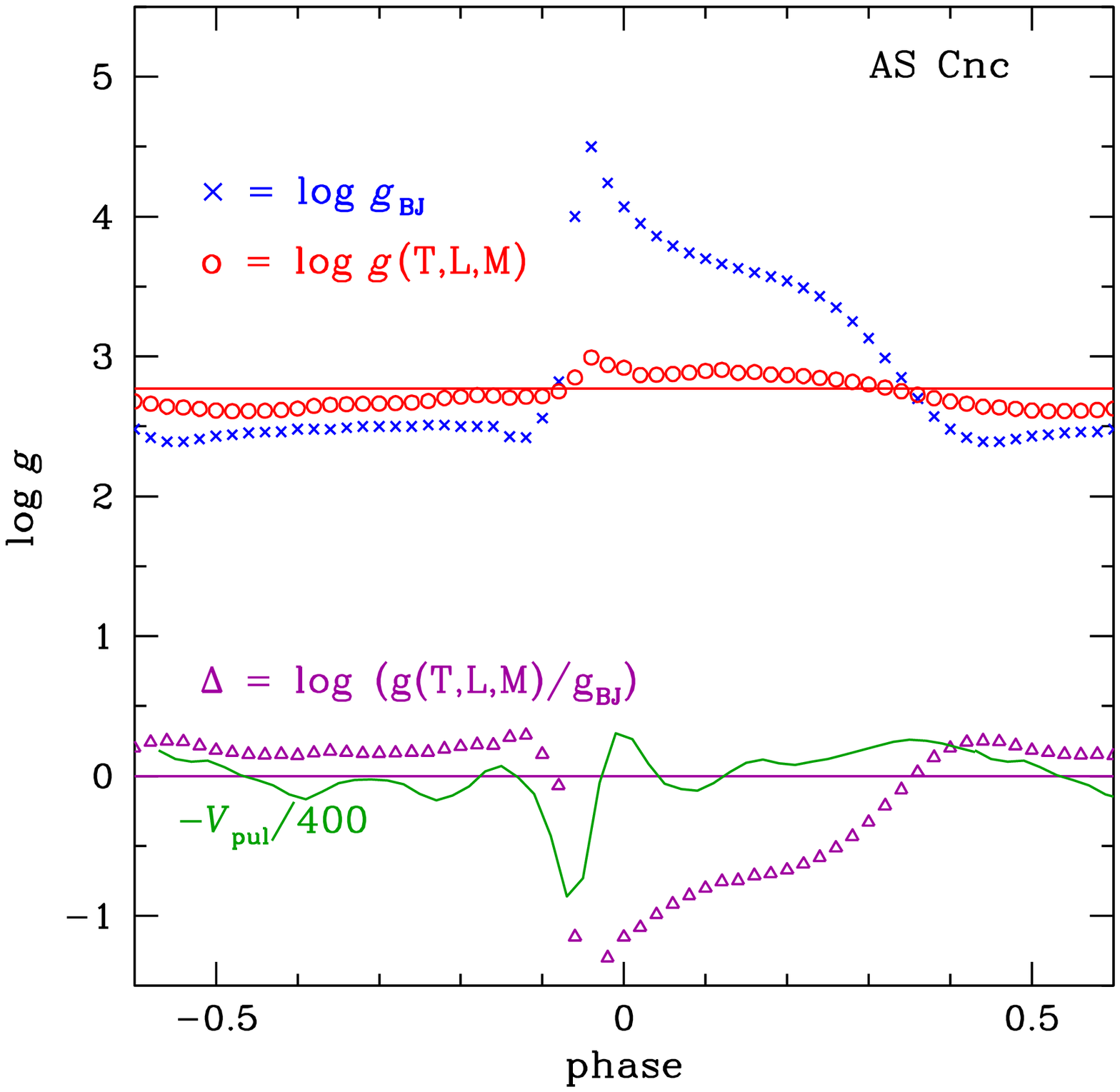}
\includegraphics{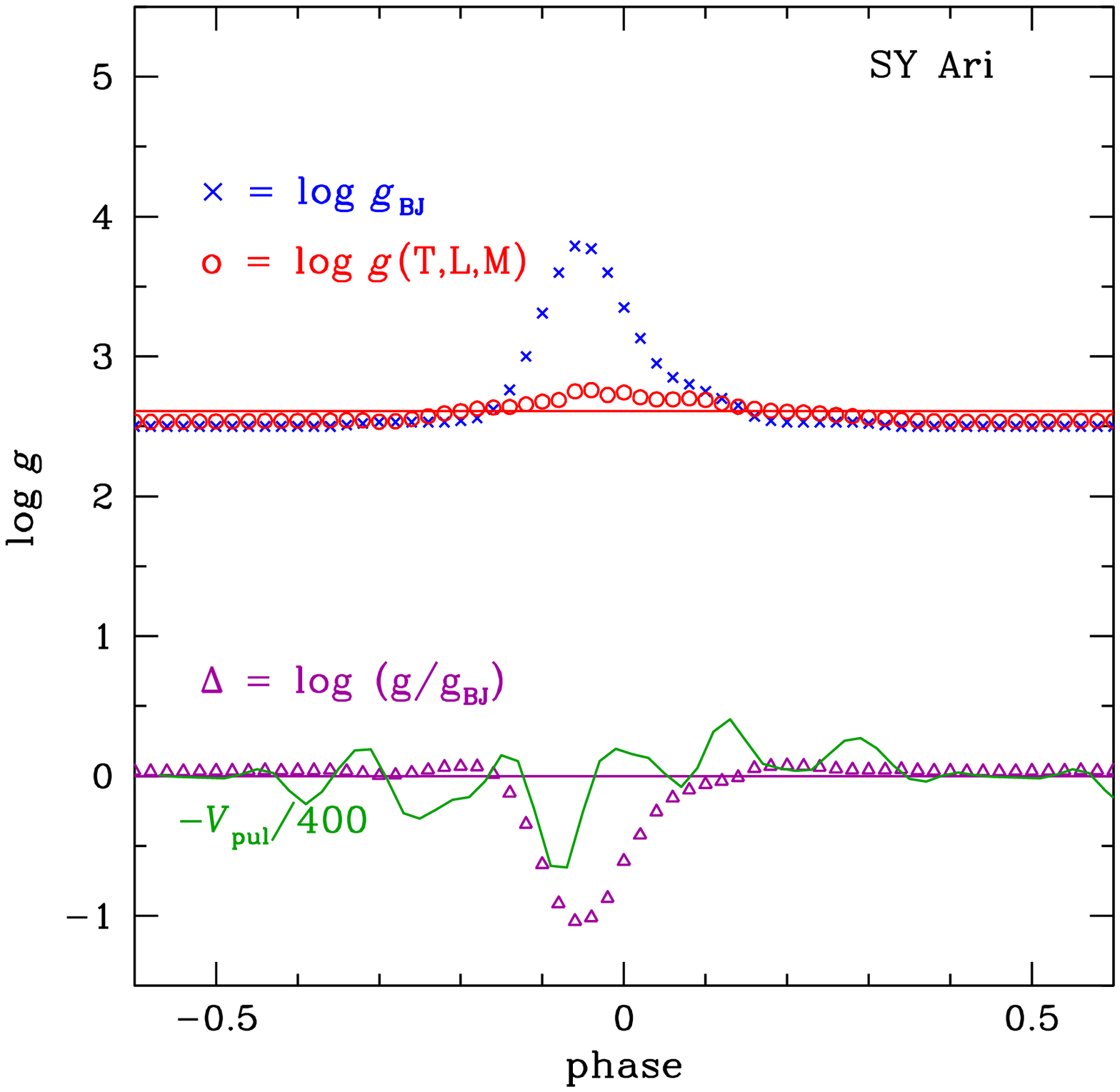}
\includegraphics{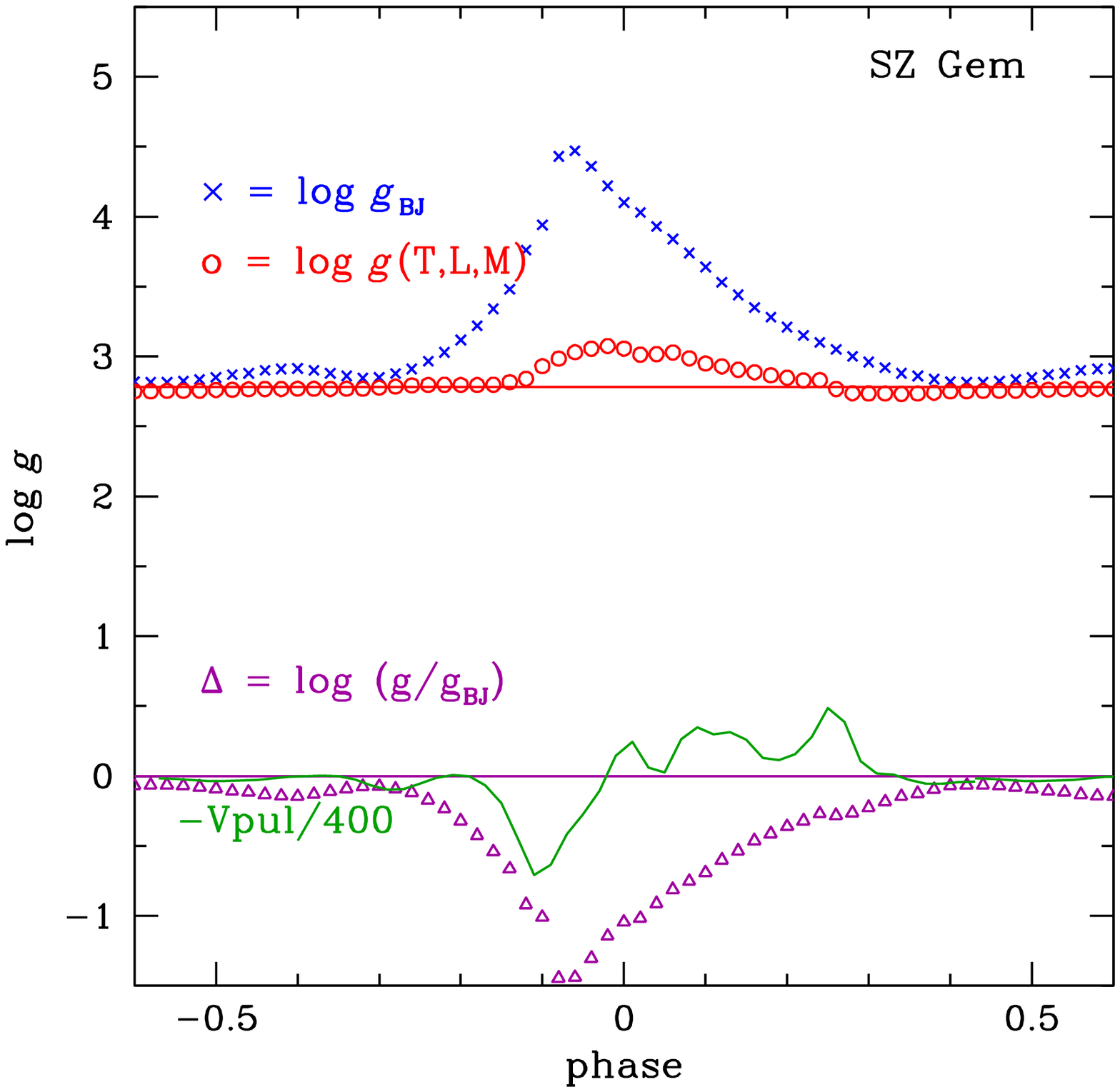}
\includegraphics{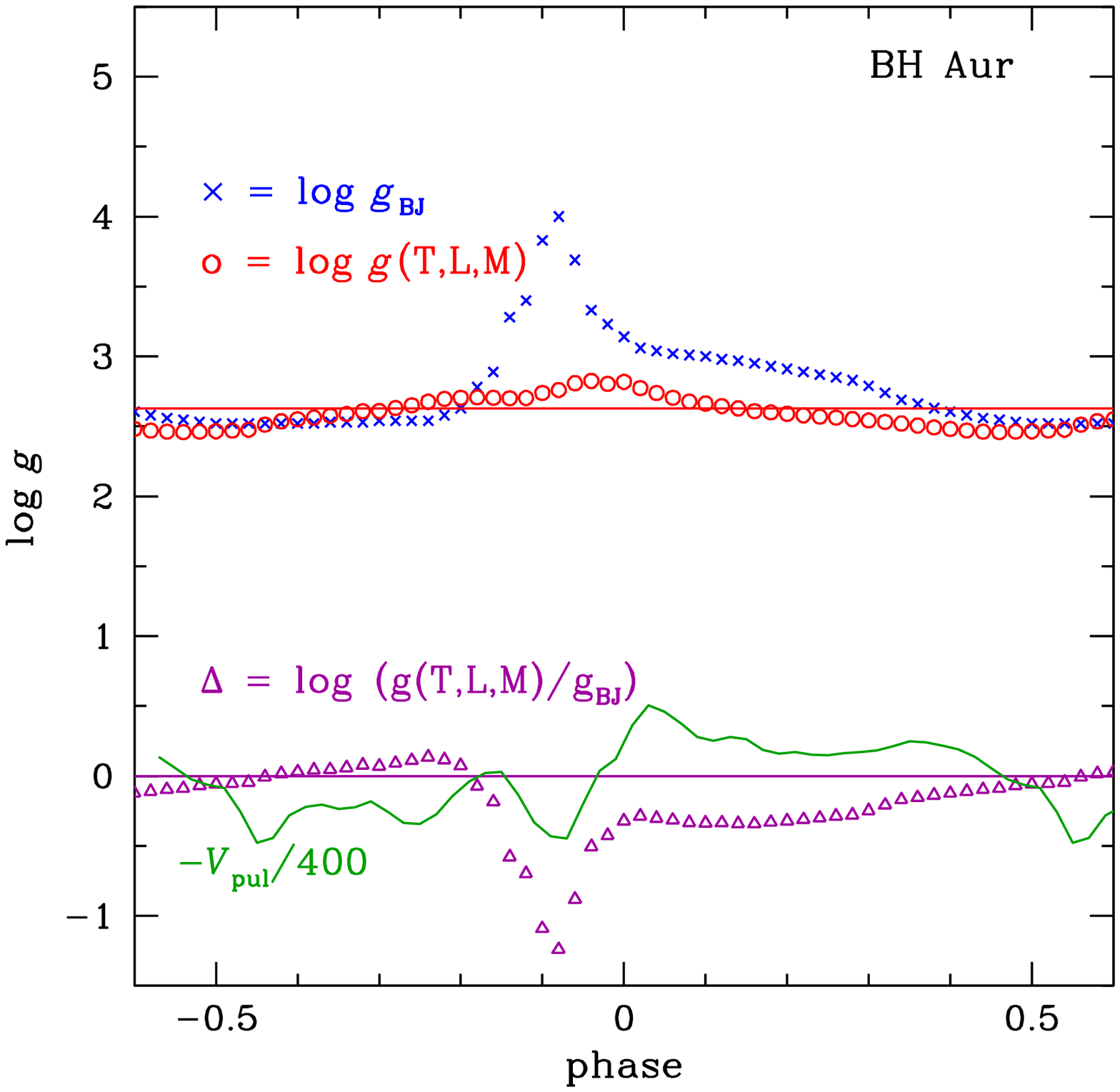}
}
\caption[logg curves]
{Run of the two aspects of $\log g$\ 
(resampled data, see Sect.\,\ref{resampling}) 
for the well observed stars. 
{\bf Top curves}: The slowly varying curve is $\log g(T,L,M)$\ derived from 
Eqs.\,\ref{elogLRT} and \ref{elogMRg} 
(thus using $T$ and $L$ derived from $y$,$b-y$), 
and the full horizontal line gives the average of $ \langle g(T,L,M) \rangle$ 
over the cycle (in logarithmic form; see Table\,\ref{ttabaver}). 
The peaked curve shows the gravity, $\log g_{\rm BJ}$, 
derived from the $b-y,c_1$ photometry and the grid calibrated 
to obtain $T_{\rm eff}$ and $\log g_{\rm BJ}$. 
{\bf Bottom curves}: 
the ratio $g(T,L,M)/g_{\rm BJ}$ is shown on a logarithmic scale. 
Clearly, when $\Delta = \log (g(T,L,M)/g_{\rm BJ}) < 0$ 
the atmosphere gas must be quite compressed. 
The curve of $V_{\rm pul}$ has added  
(in units as in Fig.\,\ref{favvrad}). 
}
\label{glog}
\end{figure*}

The difference between our velocity curves and those of Liu \& Janes 
is mostly caused by the drastic reduction of $R$ we find at $\Phi\simeq 0.9$. 
Eliminating this behaviour from our $R$ curves 
(eliminating the $R$ data for $0.85 < \Phi < 0.05$) results in 
a velocity amplitude of $V_{\rm pul}$ of up to $\simeq$90 km\,s$^{-1}$, 
quite similar to the one of Liu \& Janes mentioned above. 
In short, the radius change of an RR\,Lyrae star based on the Baade-Wesselink 
radial velocity curves does {\sl not} chime with the pronounced change in $R$ 
we derived (using Eq.\,\ref{elogLRT}) from the photometry. 
This difference is almost certainly caused by effects of optical depth. 

\subsection{Effects of optical depth $\tau$}
\label{sectau}

When analysing our data we tacitly assumed that the detected light originates, 
throughout the cycle, 
in the same gas and thus from the same atmospheric layer. 
Such assumptions have been made in almost all analyses of variable star data 
(also with the Baade-Wesselink method), but they need not be correct. 
Furthermore, Abt (1959) noted that 
the continuum light originates in layers with $\tau \simeq 0.7$ 
while the spectral lines rather are formed in layers with $\tau \simeq 0.3$.

During the cycle, compaction of the atmosphere and/or changes in 
level of ionization may lead to changes in $\tau$ in that gas, 
thus to different layers being sampled in the photometry. 
This is also the case for spectral lines: 
changes in the density and temperature of a layer due to vertical motion 
will lead to changes in the local ionization balance and the excitation state 
of ions, thus to a varying strength of the relevant absorption lines. 
During the cycle a spectral line may therefore form in different gas layers. 
Furthermore, the strength of a line formed at greater depths is also 
influenced by the level of photon scattering filling in the spectral line. 

The phase lag between lines from metals and hydrogen discovered by 
van Hoof \& Struve (1953) in a $\beta$\,Cepheid star 
has been attributed to similar optical depth effects. 
For RR\,Lyrae stars, this ``van Hoof effect'' was studied by, e.g., 
Mathias et\,al. (1995) and Chadid \& Gillet (1998). 
Our sketchy spectral data (Fig.\,\ref{rrgemspec}, Ca\,{\sc ii} and H$\gamma$) 
exhibit a phase lag too, 
which can easily be attributed to changes in the gas conditions 
that also cause the hysteresis in $T_{\rm eff}$ and $\log g_{\rm BJ}$.

In spectra of several RR\,Lyrae stars, line {\sl doubling} has been reported 
near $\Phi=0$ (see, e.g., Sanford 1949). 
Considerable increases in line width may also occur near that phase 
(e.g., Fe\,{\sc ii} lines; see, e.g., Chadid 2000). 
Spectral line doubling implies that there are gas layers in the same line of 
sight with the same ions but which are at different (radial) velocities. 
This doubling is seen mostly in the lower level Balmer series lines 
but also in other intrinsically strong lines such as Ca H\&K. 
Thus, the spectral absorption in those lines comes from 
different geometric depths in the atmosphere. 
Note that when for the determination of radial velocities template spectra 
are used, any velocity differences between ionic species are averaged out. 

\subsection{Photometry and spectroscopy sample different layers}

Consider an outward moving dense layer of the stellar atmosphere. 
Upon its expansion and cooling, its continuum optical depth becomes smaller. 
Thus, the level of $\tau\simeq 0.7$ from which the continuum light emerges 
moves physically downward through the gas, 
leading to smaller photometrically derived $R$ values. 
If this intrinsic $\tau$-level moves rapidly, 
the concomittant rapid decrease in derived $R$ 
is naturally interpreted as an extremely rapid downward velocity, 
even when the gas itself hardly has a vertical motion. 
The level from which the (metal) spectral lines emerge 
(near $\tau \simeq 0.3$) 
will exhibit this effect later, i.e., 
only after the gas has been lifted further 
to larger radii (cooled and rarefied further to lower\,$\tau$). 
This happens later than the change in the gas levels 
releasing the continuum light. 
In this case, geometrically deeper levels may then be 
(spectroscopically) sampled suggesting a decrease in $R$ 
even if the gas itself still moves outward.  
We note one can see a difference in that sense between lines of 
Fe\,{\sc ii} and Fe\,{\sc i} (Chadid \& Gillet 1998). 
Finally, the intrinsically strong lines, 
such as the lower Balmer series lines and a few other lines with 
large intrinsic absorption capability, will exhibit this effect yet later. 
However, in deeper layers (at different radial velocity) 
the local Balmer absorption may already show up with its own radial velocity 
($\lambda$-shifted with respect to the Balmer line in higher layers), 
thus leading to line-doubling from lower levels of the Balmer series. 

The presented data do not allow us to asses 
the details of these optical depth effects. 
We emphasize that the parameters derived from the photometry 
always refer to the conditions in the layer with $\tau \simeq 0.7$, 
the layer from which the measured continuum light emanates. 
It may be that some of our interpretations need to be revised once 
more detailed measurements of the particular gas layers would be available. 

The differences between our run of $R$ (and of $V_{\rm pul}$ derived from it) 
and the Baade-Wesselink run of $V_{\rm rad}$ (and of $R$ derived from it) can 
be understood as being caused by the sampling of layers with different $\tau$. 
The Baade-Wesselink method samples layers with $\tau$ smaller 
than $\tau$ of the layers sampled with continuum photometry. 
It thus is unsurprising that the run of $R$ 
as derived from these methods differs in important ways, 
all due to the physical condition ($\tau$) of the gas layer from which 
the measured radiation (be it spectral lines or the continuum) emerges. 

At the end of Sect.\,\ref{avvrad} it was noted that oscillations 
are visible in $V_{\rm pul}$ with $P/7$. 
Perhaps these variations are also caused by temporal small 
optical depth effects ($T_{\rm eff}$ and $\log g_{\rm BJ}$) 
and do not represent fluctuations in gas velocity.

\section{The variations in $\log g(T,L,M)$\ and $\log g_{\rm BJ}$, and verifications of distance and mass}
\label{secggeff}
\label{distmass}

The run of $\log g(T,L,M)$ was derived from Eq.\,\ref{elogMTgL}. 
The actual surface gravities, $\log g_{\rm BJ}$, were derived from $b-y,c_1$. 
In Fig.\,\ref{glog} we plot three parameters: $\log g(T,L,M)$, 
$\log g_{\rm BJ}$ and the logarithmic ratio $\log (g(T,L,M)/g_{\rm BJ})$. 
When $\log g(T,L,M) - \log g_{\rm BJ} <0$ (or $\log (g(T,L,M)/g_{\rm BJ}) <1$) 
the atmosphere is clearly compressed (collapsed). 

The data in Fig.\,\ref{glog} indicate that in the quiet part of the cycle 
($0.4<\Phi<0.8$) the gravity $\log g(T,L,M)$ is, for several stars, 
not equal to $\log g_{\rm BJ}$. 
However, in those quiet parts of the cycle, 
all parameters are apparently stable 
(there are most likely no optical depth effects at this point) and 
so we would expect $\log g(T,L,M)$ and $\log g_{\rm BJ}$ to be equal there. 
We can explore how one might adjust $d$ and $M$ to make $\log g(T,L,M)$ from 
Eq.\,\ref{elogMTgL} equal to $\log g_{\rm BJ}$ in that phase interval.
For RR\,Lyrae stars, 
that phase ($\Phi\simeq 0.6$) has $T_{\rm eff} \simeq 6000$~K, 
only a little higher than $(T_{\rm eff})_{\odot}$. 

The distances of our stars were defined as described in Sect.\,\ref{secobs}. 
They are taken from the literature and they 
are primarily determined from an adopted reference value of $M_V$. 
For some stars we calculated $d$ from $M_V$ (including effects of [Fe/H]). 

We recall from Sect.\,\ref{errorbudget} (on the error budget) that 
the sum of the measurement errors in $\log g(T,L,M)$ and $\log g_{\rm BJ}$ 
may be as large as 0.08 dex. 
If these two gravities were to be equal
and if we assumed the full margin of error of 0.08, then, e.g.,  
$\log g(T,L,M)$ for RR Gem should be reduced by $\simeq$0.22 dex, 
that of AS Cnc by $\simeq$0.12, 
while that of TW Lyn should be larger by $\simeq$0.07 dex. 

If the adopted $M_V$ (or $d$) and $M$ of our stars were wrong 
and had to be changed, 
this would produce (when using Eq.\,\ref{elogMTgL}) 
the tabulated effects: 
\begin{center}
\begin{tabular}{ccc|ccc}
\hline
\multicolumn{3}{c}{assumed change} & \multicolumn{3}{c}{effects} \\
\hline
 $M_V$ & ($d$) & $M$ & $\log g(T,L,M)$ & $L$ & $R$\\
\hline
 $-0.20$ & $+10$\% & & $-0.04$ & $+20$\% & +10\%\\
      &       & +10\% & +0.04 & - & - \\
\hline
\end{tabular}
\end{center}
\noindent
We note again that changing the distance is for the stars observed 
only implicit; 
distances of RR\,Lyrae stars have (except for those 
in the papers referred to in Sect.\,\ref{parpulsating}) 
not been determined by other means than through $M_V$. 
Thus only $M_V$ or the mass $M$ can be changed. 

For our stars a change in either $M_V$ or $M$ 
(or perhaps a combination thereof) 
is needed to make $\log g(T,L,M)$ and $\log g_{\rm BJ}$ match. 
The changes in $M_V$ indicated below are within the range permissible by the 
non-uniqueness of $M_V$, as mentioned in Sect.\,\ref{errorbudget}. 
For the stars shown in Fig.\,\ref{glog} the changes are: \\
\noindent
- \object{RR Gem}: $M_V$ 1.0 mag brighter, $M$ 50\% smaller. 
  The optimum would be to change $M_V$ 
  leading to $L$ larger by a factor 2.5 (0.4 dex in $\log L$). 
  Reducing the mass by 50\% is no option, it then would be too low. 
  A combination of changes might also fit.\\
- \object{TW Lyn}: $M_V$ 0.4 mag fainter, $M$ 20\% larger. 
  Changing $M_V$ is not the right option because its $L$ value 
  would, in Fig.\,\ref{tefflloop}, be even further below the ZAHB. 
  Hence $M$ should be 20\% higher at 0.85 M$_{\odot}$. 
  Note that (in Fig.\,\ref{tefflloop}) 
  the mean value of $L$ can be made larger 
  when adopting a yet larger mass.\\
- \object{AS Cnc}: $M_V$ 0.6 mag brighter, $M$ 30\% lower. 
  Changes similar to but smaller than those for RR Gem would be needed.\\
- \object{SY Ari}: no changes are needed within the errors.\\
- \object{SZ Gem}: $M$ smaller by 0.1 dex, to give $M= 0.56$ M$_{\odot}$.\\
- \object{BH Aur}: no changes are needed within the errors. \\
- \object{TZ Aur}: to make the two gravities match, 
$\log g(T,L,M)$ should be $\simeq$0.3 dex lower. 
This would be achieved if $M$ were considerably lower. 
Alternatively, $M_V$ should (with $M=0.7$\,M$_{\odot}$) be so much brighter, 
that $L$ is doubled. 
In both cases, TZ\,Aur would then be a well-evolved HB star. 

We do not comment on the less well-observed stars 
(the bottom four in Table\,\ref{ttabaver}).

\section{The lightcurve bump}

\begin{figure}
\resizebox{\hsize}{!}
{\includegraphics{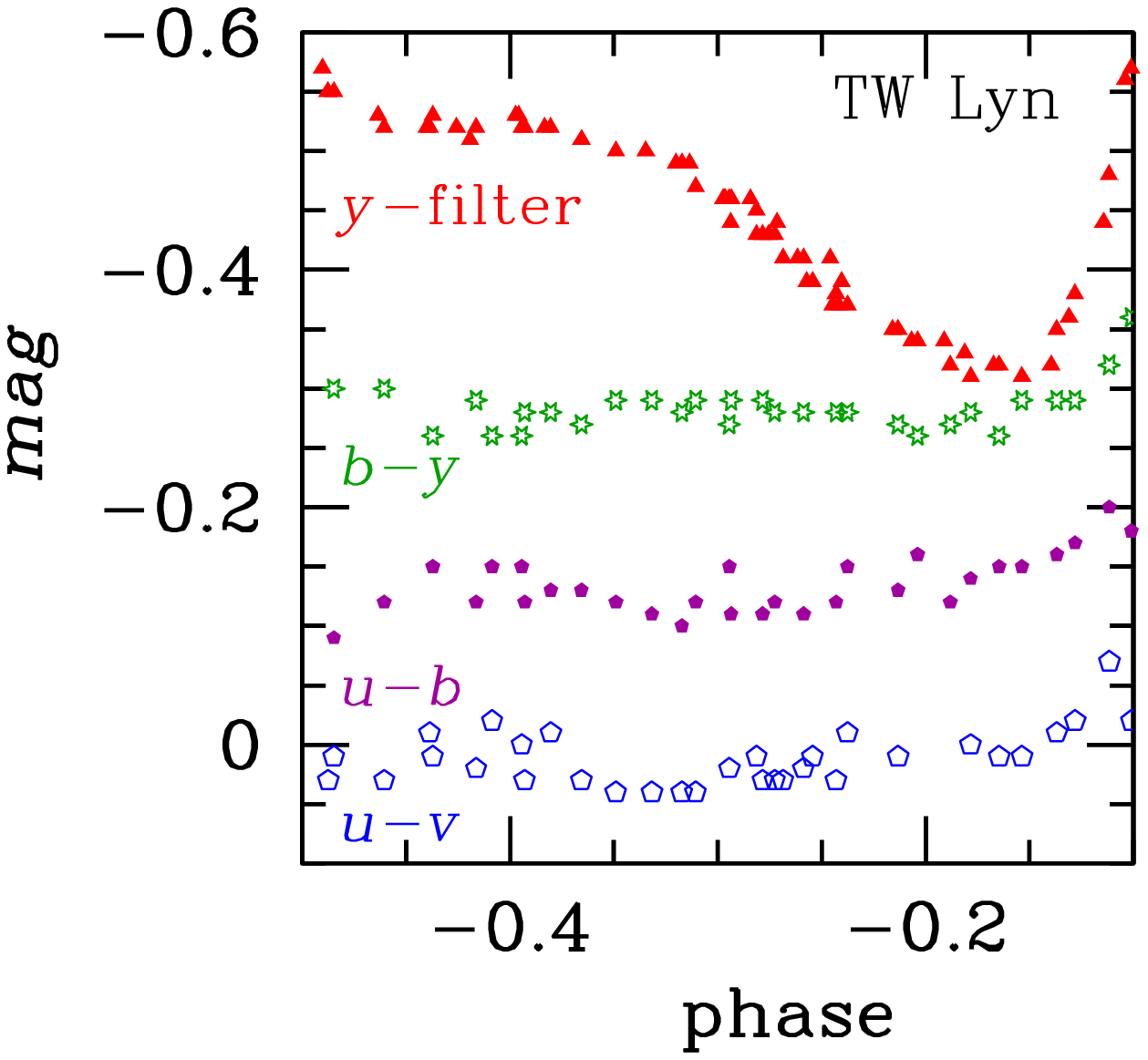}\includegraphics{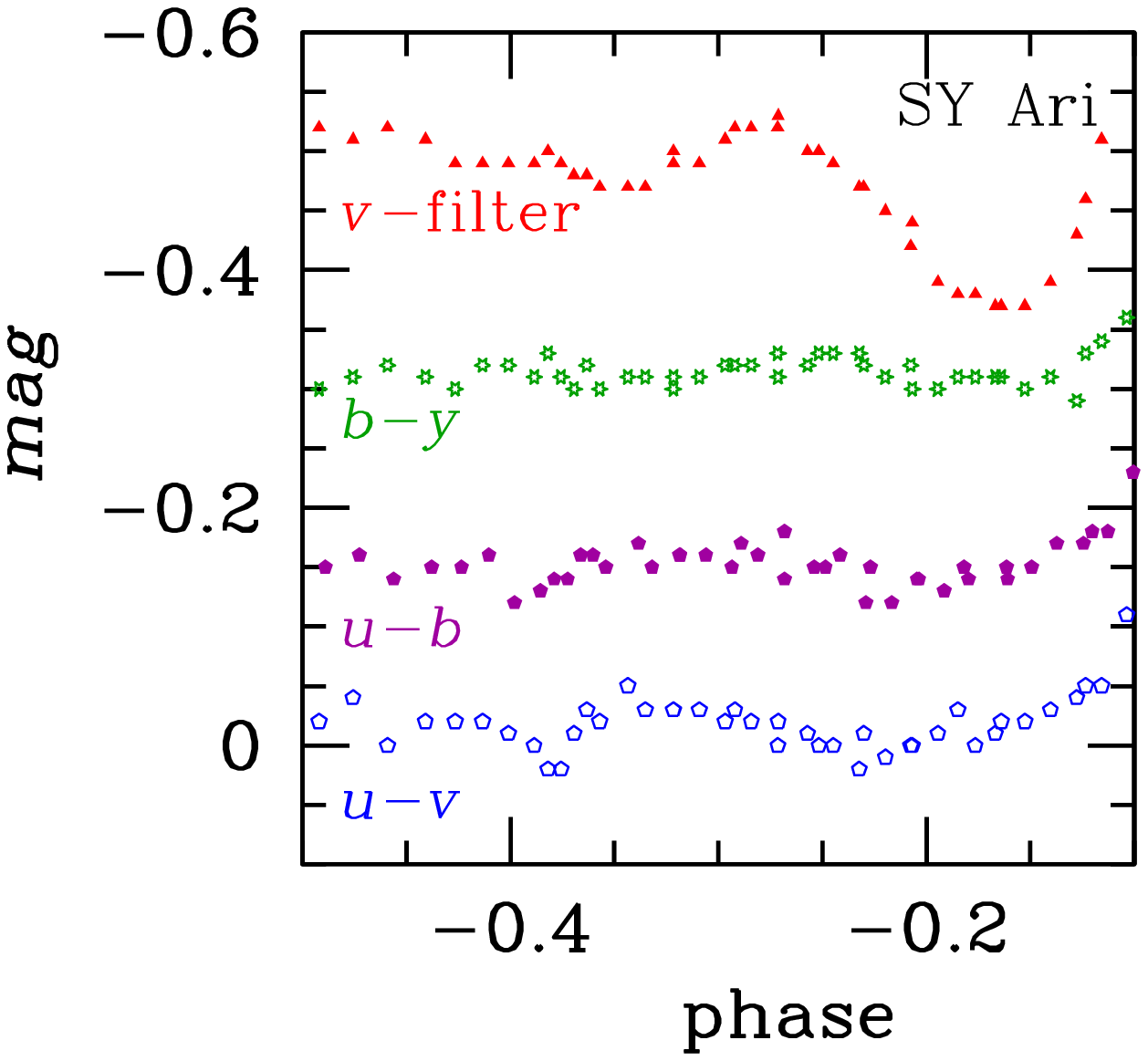}\includegraphics{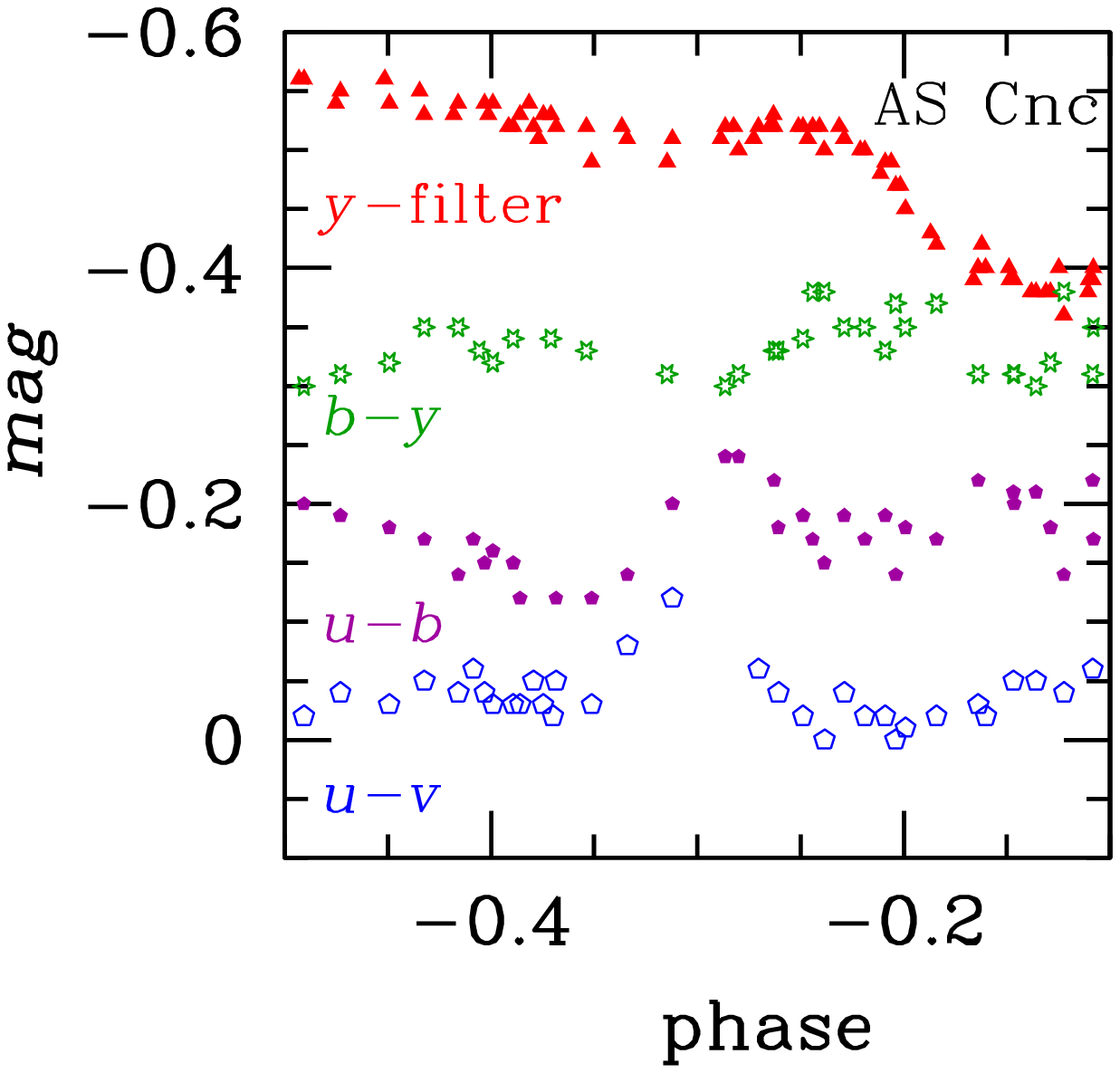}}
\caption[]{
Three examples of Str\"omgren-photometry light curves of RR\,Lyrae stars 
near the possible ``bump'' in brightness. 
\object{TW Lyn} either does not show a bump or its bump is rather extended 
in phase; 
there is no or little effect on the colour indices. 
\object{SY Ari} does exhibit a bump in $y$ 
but there are no effects in its colour indices. 
For \object{AS Cnc}, the colour indices $u-b$ and $u-v$ clearly change 
before $y$ changes, 
indicating a change in the Balmer jump before a brightness change. 
The index $b-y$ (representing temperature) changes somewhat later.
}
\label{bump}
\end{figure}

The stars in our programme were primarily selected from the class said to have 
lightcurve ``bumps''. 
These bumps in the run of brightness need not be very pronounced 
(for $y$-magnitudes see Fig.\,\ref{flightcurves}). 
One goal of the observations was to investigate 
whether these bumps in brightness affect the colours of the stars. 
Examples of the bump region for a few stars are given in Fig.\,\ref{bump}. 

First, we note that most of our stars with good light-curve coverage 
hardly show the light-curve bump in the Str\"omgren colour indices. 
Among these are \object{TW Lyn}, \object{BH Aur}, \object{X CMi}, 
\object{TZ Aur}, \object{BR Tau}, and \object{RR Gem}, which 
have either a prolonged low level bump or no clear colour-index bump at all. 
While \object{SY Ari} has a bump in $y$ that has no effect 
in its colour indices, \object{AS Cnc} does show an effect. 
For the other stars, light-curve coverage was either poor or missing in this 
range of phases. 

In \object{AS Cnc} the brightening and dimming is clearly recognisable 
and takes place in a short timespan. 
We note that the colour indices $u-b$ and $u-v$ change 
before $y$ brightens. 
Both indices become bluer meaning that the Balmer continuum radiation 
brightens without an increase in $T_{\rm eff}$. 
In this part of the cycle, the atmosphere is cool and 
the Balmer continuum brightening could mean that the gas cools even further 
leading to an additional reduction in opacity 
based on the lower Balmer level excitation. 
Since $T$ has not yet changed, 
the changes should be due to a reduction in gas density. 
Does extra light escape because of the lower $\tau$? 
At this point in the cycle, the atmosphere is quite extended 
and about to shrink. 
This happens at $\Phi \simeq -0.1$ (see Fig.\,\ref{bump} and earlier ones). 

Bono \& Stellingwerf (1994) attribute the bump feature to shockfronts. 

\section{Further remarks} 

\subsection{Comparison with models} 

Numerous models have been constructed for the pulsation 
and for various other detailed aspects of RR\,Lyrae 
(see, e.g., Fokin et\,al. 1999). 
A comparison with models is of limited value since 
most models are one-dimensional. 
More importantly, 
it is not always clear how theory has been transformed to observables. 
Do the theoretical predictions for $V$, $R$, spectral line strengths, 
and velocities all refer to information from the layer 
with the relevant $\tau$ (thus to what one really would observe) 
or to a matter-defined gas layer? 

Parameters of RR\,Lyrae stars have been determined 
using different kinds of data and with different modelling. 
We refer to the analysis of S\'odor et\,al. (2009) for results 
based on the Baade-Wesselink method and its refinements 
as well as to additional literature. 

\subsection{Blazhko effect}

A good portion of RR\,Lyrae stars shows cycle-to-cycle variations 
in their light curves, the so-called Blazhko effect 
(for references see, e.g., Jurcsik et\,al. 2009). 
Some of the stars of our sample are also known to be Blazhko stars 
but they exhibit these variations only at a moderate level. 
However, these variations may affect the stellar parameters derived. 
The data we acquired cover to short a time span to address and assess 
possible cycle-to-cycle variations in the stellar parameters. 

\subsection{Relation between stellar mass and kinematics}

Six of our stars were part of the study of kinematics of RR\,Lyrae stars 
(Maintz \& de Boer 2005). 
Of these, five (\object{CI And}, \object{AR Per}, \object{TZ Aur}, 
\object{RR Gem}, and \object{TW Lyn}) 
are stars of the disk population (according to their kinematics), 
while \object{SZ Gem} is a star with halo kinematics. 
Halo RR\,Lyrae are understood to be older than disk RR\,Lyrae, 
so should (on average) have a lower mass. 

For \object{TW Lyn}, we found that its mass should be higher than the 
adopted reference value and instead be $\simeq$0.85 M$_{\odot}$, 
which chimes with it being a star of the disc. 
We note that S\'odor et\,al. (2009) found 
a mass of $\simeq$0.82 M$_{\rm \odot}$ for \object{RR Gem}, 
our analysis indicated $\simeq$0.7 M$_{\odot}$. 
For \object{SZ Gem}, we found a mass of $\simeq$0.56\,M$_{\odot}$, 
this low mass being in line with the expected 
(larger age and) lower mass of halo stars. 

\object{BH Aur} and \object{SY Ari} are found to correspond to the reference 
value of $M=0.7$ M$_{\odot}$, which is in line with their being disk stars.

\section{Conclusions}

Based on simultaneous Str\"omgren photometry and a comparison 
of the photometric indices with a calibrated $T_{\rm eff}$, $\log g$ grid, 
accurate atmospheric parameters of RR\,Lyrae stars have been determined. 
Curves of phase related runs of $\log g_{\rm BJ}$ 
(the gravity from the Balmer jump) and $T_{\rm eff}$ were derived. 
One can then calculate the change in radius of the stars over the cycle. 
By including additionally obtained spectra, one arrives at the following 
description of the behaviour of RR\,Lyrae stars.

The straightforward interpretation of the data derived for $R$ is that 
at phase $\Phi \simeq 0.9$ the atmosphere begins to collapse 
soon reaching a high gas density 
(when the star has its greatest brightness). 
This manifests itself in a large Balmer jump (large $\log g_{\rm BJ}$). 
Within an interval $\Delta\Phi\simeq 0.1$ the atmosphere then expands again. 

An alternative interpretation is that at phase $\Phi \simeq 0.9$ 
the optical depth of the extended outer layers has become so small 
that the level of $\tau \simeq 0.7$ sampled in the photometry 
rushes through the gas inward mimicking a drastic reduction in radius. 
The contraction taking place anyway then elevates the density in that level 
having $\tau \simeq 0.7$, raising the optical depth there so that 
photometry subsequently samples layers again higher up in the atmosphere, 
suggesting a rapid expansion. 
The sudden increase in density manifests itself 
in lage values of $\log g_{\rm BJ}$. 

During all these changes the atmosphere exhibits oscillation ripples 
with a rhythm of $P/7$. 

A possible mismatch of the observed $\log g_{\rm BJ}$ and the calculated 
$\log g(T,L,M)$ in the quiet, descending part of the light curve can be used 
to assess the applicability of the mean parameters adopted for the stars, 
i.e., the absolute brightness $M_V$ (no geometric distances are known) 
and the mass~$M$. 
One can then determine the individual values of $M_V$ and $M$ for a star. 

Extending this kind of simultaneous Str\"omgren-photometry 
while avoiding multiplexing two stars 
would provide a much denser coverage of the light curves 
and thus a more accurate determination of the cycle variation in $R$. 
If spectra were then taken simultaneously, 
in which a range of spectral lines is included 
(lower and higher ionisation stages, 
strong and weak lines of the Balmer series) 
and of a nature to allow velocity determinations, 
one may hope to distinghuish the effects caused by the sampling 
of light from layers at different optical depth.

\acknowledgements{We thank Oliver Cordes, Klaus Reif and the 
AIfA electronics group for their dedication to {B\sc usca}, 
and the staff at the Calar Alto Observatory 
and the Observatorium Hoher List for their technical support.
We thank K. Kolenberg, M. Papar\'o and K. Werner for advice. 
We are grateful that the referee, Dr. \'A. S\'odor, 
graciously gave many suggestions for improvement 
and asked pertinent questions. 
These stimulated us to carry the interpretation a step further. 
We thank C. Halliday for linguistic advice.}

%

%

\begin{thebibliography}{}
\bibitem[]{}Abt, H. 1959, ApJ, 130, 824
\bibitem[]{}Baade, W. 1926, AN, 228, 359
\bibitem[]{}Barcza, S. 2003, A\&A, 403, 683
\bibitem[]{}Beers, T. C., Chiba, M., Joshii, Y., et\,al. 2000, AJ, 119, 2688
\bibitem[]{}Bono, G., \& Stellingwerf, R. F. 1994, ApJS, 93, 233
\bibitem[]{}Breger, M. 1974, ApJ, 192, 75
\bibitem[]{}Chadid, M. 2000, A\&A, 359, 991
\bibitem[]{}Chadid, M., \& Gillet, D. 1998, A\&A, 335, 255
\bibitem[]{}Clem, J. L., VandenBerg, D. A., Grundahl, F., \& Bell, R. A. 2004, AJ, 127, 1227
\bibitem[]{}Cox, J. P. 1980, ``Theory of Stellar Pulsation'', Princeton Univ. Press
\bibitem[]{}Danziger, I. J., \& Oke, J. B. 1967, ApJ, 147, 151
\bibitem[]{}de Boer, K. S., Schmidt, J. H. K., \& Heber, U. 1995, A\&A, 303, 95
\bibitem[]{}de Boer, K. S., \& Seggewiss, W. 2008, ``Stars and Stellar Evolution''; EDP Sciences
\bibitem[]{}de Boer, K. S., Tucholke, H.-J., \& Schmidt, J. H. K. 1997, A\&A, 317, L23
\bibitem[]{}Fernley, J. 1994, A\&A, 284, L16
\bibitem[]{}Fernley, J., Barnes, T. G., Skillen, I., Hawley, S. L., 
Hanley, C. J., Evans, D. W., Solano, E., \& Garrido, R. 1998, A\&A, 330, 515
\bibitem[]{}Fokin, A. B., Gillet, D., \& Chadid, M. 1999, A\&A, 344, 930
\bibitem[]{}Gautschy, A., \& Saio, H. 1995, ARAA, 33, 75
\bibitem[]{}Gratton, R. F. 1998, MNRAS, 296, 739
\bibitem[]{}Hilker, M. 2000, A\&A, 355, 994
\bibitem[]{}Jurcsik, J., S\'odor, \'A., Szeidl, B., et\,al. 2009, MNRAS, 400, 1006
\bibitem[]{}Layden, A. C. 1994, AJ, 108, 1016
\bibitem[]{}Ledoux, P., \& Walraven, Th. 1958, in Handbook of Physics, Vol LI, p.353
\bibitem[]{}Liu, T., \& Janes, K. A. 1989, ApJS, 69, 593 
\bibitem[]{}Liu, T., \& Janes, K. A. 1990, ApJ, 354, 273 
\bibitem[]{}Lub, J. 1977a, PhD Thesis, Univ. Leiden
\bibitem[]{}Lub, J. 1977b, A\&AS, 29, 345 
\bibitem[]{}Lub, J. 1979, AJ, 84, 383
\bibitem[]{}Maintz, G. 2005, A\&A, 442, 381
\bibitem[]{}Maintz, G. 2008a, PhD Thesis, Univ. Bonn
\bibitem[]{}Maintz, G. 2008b, BAVSR, 57, 150
\bibitem[]{}Maintz, G., \& de Boer, K. S. 2005, A\&A, 442, 229
\bibitem[]{}Mathias, P., Gillet, D., Fokin, A. B., \& Chadid, M. 1995, A\&A, 298, 843
\bibitem[]{}McNamara, D. H., \& Feltz, K. A. 1977, PASP, 89, 699
\bibitem[]{}Moehler, S., Heber, U., \& de Boer, K. S. 1995, A\&A, 294, 65
\bibitem[]{}Moehler, S., Heber, U., \& Rupprecht, G. 1997, A\&A, 319, 109
\bibitem[]{}Oke, J. B. 1966, ApJ, 145, 468
\bibitem[]{}Oke, J. B., \& Bonsack, S. J. 1960, ApJ, 132, 417
\bibitem[]{}Oke, J. B., Giver, L. P., \& Searle, L. 1962, ApJ, 136, 393
\bibitem[]{}Papar\'o, M., Szab\'o, R., Benk\'o, J. M., et al. 2009, 
in ``Stellar Pulsation: Challenges for Theory and Observation'', J.A.Guzik 
and P. Bradley (eds.), AIP Conf. Proc. 1170, p. 240
\bibitem[]{}Perry, C. L., Olsen, E. H., \& Crawford, D. L. 1987, PASP, 99, 1184
\bibitem[]{}Preston, G. W., \& Paczynski, B. 1964, ApJ, 140, 181
\bibitem[]{}Reif, K., Bagschik, K., de Boer, K. S., Schmoll, J., 
M\"uller, Ph., Poschmann, H., Klink, G., Kohley, R., Heber, U., \& Mebold, U. 
1999, SPIE Vol. 3649, 109 
\bibitem[]{}Renzini, A., Greggio, L., Ritossa, C., \& Ferrario, L. 
1992, ApJ, 400, 280
\bibitem[]{}Sandage, A., \& Tamman, G. A. 2006, ARAA, 44, 93
\bibitem[]{}Sandford, R. F. 1949, ApJ, 109, 208
\bibitem[]{}Schmidt-Kaler, T. 1982, Landolt B\"ornstein; 
Mathematics for Engineers; X.~Physics and Applied Physics for Engineers; 
Astronomy/Astrophysics
\bibitem[]{}Siegel, M. J. 1982, PASP, 94, 122
\bibitem[]{}S\'odor, \'A., Jurcsik, J., \& Szeidl, B. 2009, MNRAS, 394, 261
\bibitem[]{}Struve, O. 1947, PASP, 59, 192
\bibitem[]{}Tsujimoto, T., Miyamoto, M., \& Yoshi, Y. 1998, ApJ, 492, L79
\bibitem[]{}van Albada, T. S., \& de Boer, K. S. 1975, A\&A, 39, 83 
\bibitem[]{}van Hoof, A., \& Struve, O. 1953, PASP, 384, 158
\bibitem[]{}Wesselink, A. J. 1946, Bull.\,Astr.\,Inst.\,Neth., 10, 83
\end{thebibliography}
\end{document}